\documentclass[12pt]{article}
\usepackage{mathrsfs}
\usepackage{amssymb}
\usepackage{dsfont}
\usepackage{amsmath}
\usepackage{graphicx}
\usepackage{subfigure}
\usepackage{etex}
\usepackage{booktabs}

\normalsize
\input{epsf}

\makeatletter

\newcommand{\Rmnum}[1]{\expandafter\@slowromancap\romannumeral #1@}
\makeatother

\begin{document}
\addtolength{\baselineskip}{.20mm}
\newlength{\extraspace}
\setlength{\extraspace}{2mm}
\newlength{\extraspaces}
\setlength{\extraspaces}{2mm}

\newcommand{\newsection}[1]{
\vspace{15mm} \pagebreak[3] \addtocounter{section}{1}
\setcounter{subsection}{0} \setcounter{footnote}{0}
\noindent {\Large\bf \thesection. #1} \nopagebreak
\medskip
\nopagebreak}
\newcommand{\newsubsection}[1]{
\vspace{1cm} \pagebreak[3] \addtocounter{subsection}{1}
\addcontentsline{toc}{subsection}{\protect
\numberline{\arabic{section}.\arabic{subsection}}{#1}}
\noindent{\large\bf 
\thesubsection. #1} \nopagebreak \vspace{3mm} \nopagebreak}
\newcommand{\ba}{\begin{eqnarray}
\addtolength{\abovedisplayskip}{\extraspaces}
\addtolength{\belowdisplayskip}{\extraspaces}

\addtolength{\belowdisplayshortskip}{\extraspace}}

\newcommand{\be}{\begin{equation}
\addtolength{\abovedisplayskip}{\extraspaces}
\addtolength{\belowdisplayskip}{\extraspaces}
\addtolength{\abovedisplayshortskip}{\extraspace}
\addtolength{\belowdisplayshortskip}{\extraspace}}
\newcommand{\ee}{\end{equation}}
\newcommand{\STr}{{\rm STr}}
\newcommand{\figuur}[3]{
\begin{figure}[t]\begin{center}
\leavevmode\hbox{\epsfxsize=#2 \epsffile{#1.eps}}\\[3mm]
\parbox{15.5cm}{\small
\it #3}
\end{center}
\end{figure}}
\newcommand{\im}{{\rm Im}}
\newcommand{\calm}{{\cal M}}
\newcommand{\call}{{\cal L}}
\newcommand{\sect}[1]{\section{#1}}
\newcommand\hi{{\rm i}}
\def\bea{\begin{eqnarray}}
\def\eea{\end{eqnarray}}

\begin{titlepage}
\begin{center}

\vspace{3.5cm}

{\Large \bf{The Cosmological  Effect of  CMB/BAO  Measurements}}\\[1.5cm]

{Yi Zhang $^{a,b,}$\footnote{Email: zhangyia@cqupt.edu.cn}} \vspace*{0.5cm}

{\it $^{a}$College of Science, Chongqing University
of Posts and Telecommunications, \\ Chongqing 400065, China\\
 $^{b}$State Key Laboratory of Theoretical Physics, Institute of Theoretical Physics,\\ Chinese Academy of Science,  \\ Beijing 100190, China}
\date{\today}
\vspace{3.5cm}

\textbf{Abstract} \vspace{5mm}

\end{center}
  In this paper,  the CMB/BAO measurements which cover the 13 redshift data in the regime $0.106 \leq z \leq 2.34$ are given out. The CMB/BAO samples are based on  the BAO distance ratios $r_{s}(z_d)/D_{V}(z)$
and the CMB acoustic scales $l_{A}$. It  could give out the accelerating behaviors of the  $\Lambda$CDM, $w$CDM and o$\Lambda$CDM models.  As the direction of the degeneracy of $\Omega_{m0}-w$  and  $\Omega_{m0}-\Omega_{k0}$ are different for the CMB/BAO and BAO data, the CMB/BAO data show ability of breaking parameter degeneracy. Our tightest constraining results is from the BAO+Planck/BAO+$\Omega_{b}h^2$+$\Omega_{m}h^2$  data which has  $\Omega_{m0}$ tension,  but doesn't have $H_{0}$ tension with the Planck result.  The extending parameters $w$  and  $\Omega_{k0}$ could alleviate the $\Omega_{m0}$ tensions slightly.

\end{titlepage}

\section{Introduction}\label{sec1}
  The  CMB/BAO data \cite{Sollerman:2009yu} is a  kind of observational data which try to  exact more information from the important cosmological experiments: CMB (Cosmic Microwave Background) and BAO  (Baryon Acoustic Oscillation). 
 The CMB  probes the rate of the expansion at redshift $z \sim 1100$ \cite{Bennett:2012zja,Ade:2013zuv} and the BAO    technique   provides a distance-redshift relation at low redshifts \cite{Percival:2007yw,Beutler:2011hx, Percival:2009xn,Kazin:2009cj,Padmanabhan:2012hf,Blake:2011en,Anderson:2013zyy,Delubac:2014aqe,Blake:2011wn,Anderson:2012sa,Busca:2012bu,Veropalumbo:2013cua,Manera:2012sc}. 
 CMB and BAO are observations to different  issues and challenges, and affected by different physics; one is by any non-standard early-universe physics, the other is by late time expansion mainly.  Typically, the CMB/BAO cosmological constraining results are discussed with other low-redshift data or the CMB data to fix tightly constraining portions of parameter space. In this letter, the CMB/BAO data  is  sufficient to permit a meaningful comparison with type IA supernovae (SNe) measurement  without strong CMB priors.

In the first CMB/BAO paper \cite{Sollerman:2009yu}, the number of CMB/BAO data   is just $2$. As the development of BAO survey,  and the number is  increased to $5$  in Refs.\cite{Bennett:2012zja,Li:2013oba,ZHANG:2013dia}.  To make clear  the constraining effects of the CMB/BAO data, here we collect 13 observational BAO data and choose the  Planck or WMAP$9$ survey\footnote{In the simulated BAO data which has $21$ data  \cite{Lazkoz:2013ija},  the $\Lambda$CDM model could be constrained  as well.}.  One benefit of the BAO data   is that it is  limited by the statistical error, rather than systematic uncertainties \cite{Weinberg:2012es}.     As the newly released CMB data,  Planck results, consist with the  results from BAOs data \cite{Addison:2013haa}. 
We concentrate on testing the  $\Lambda$CDM model and its extensions with the CMB/BAO data. 

The rationale for considering the $\Lambda$CDM model is that one could consider the cosmological constant as a ``null hypothesis" for dark energy, so it is worth exploring whether it provides a reasonable description of data.  As our dataset  larger, whether the dataset  is suitable to standard $\Lambda$CDM model is an interesting question.  
And,  a hypothetical deviation from $\Lambda$-acceleration may first appear as a tension between CMB and low-redshift data, or  arise from statistical fluctuations, systematic uncertainties that are incorrectly correctly quantified, alternatively extensions to the standard model, or some combinations of these factors. The ability to disambiguate these possibilities from current and future low-redshift experiments is crucial. So we consider  $w$CDM which adopted a   constant w (the equation of state of dark energy $p_{de}/\rho_{de}$) and o$\Lambda$CDM model which extends this model to  non-zero dimensionless energy density of curvature $\Omega_{k0}$ as well.

In this paper, we use improved CMB/BAO data to constrain the $\Lambda$CDM, $w$CDM and $\Lambda$CDM models. The rest of the paper is organized as follows.  In Section \ref{sec2},  we update the  CMB/BAO data and give out the theoretical models.
In Section \ref{sec3},  we  compare the constraining result and discuss the effects of the CMB and BAO  data. At last,  we  present a short summary in Section \ref{sec4}.

\section{ Data }\label{sec2}
The acoustic peak in the galaxy correlation function provides a standard ruler $r_{s}(z_d)/D_{V}(z) $ (or its inverse  $D_{V}(z) /r_{s}(z_d)$) which measure the distance to objects at   redshift z  of recombination in unit of the sound horizon. The sound horizon and the dubbed spherically averaged distance  are
\begin{eqnarray}
\nonumber
&&r_s(z_d)=\int_{z_d}^{\infty} \frac{c_s(z)dz}{H(z)},\\
\nonumber
&&D_{V}(z)=(\frac{(1+z)^{2}d_{A}(z_*)^{2} cz}{H})^{1/3},
\end{eqnarray}
where  H is the Hubble parameter, $c_s$ is the speed of sound before recombination,  $d_A(z_*)$  is   the co-moving angular diameter distance to recombination, $z_*$   is the redshift at recombination with the  value of $1090.48$ and $z_d$ is  the value of the redshift of the drag epoch which is gotten by  the fitting formula in Ref.\cite{Eisenstein:1997jh}.
The  observable quantities of BAO are only sensitive to the early universe physics via  the sound horizon $r_s(z_d)$. 
And, the CMB data  provide an excellent standard ruler as well which is 
  the position of the first CMB power spectrum peak which represents the angular scale of sound horizon at recombination $z_*$, dubbed the acoustic scale, 
\begin{equation}
\nonumber
l_A=\frac{\pi d_A(z_*)(1+z_*)}{r_s(z_*)}.
\end{equation}

The CMB/BAO parameter $f$ is regarded to be more suitable than the primitive CMB shift parameter $l_{A}$ for non-standard dark energy model \cite{Sollerman:2009yu,Li:2013oba,ZHANG:2013dia}:
\begin{eqnarray}
\label{f}
\nonumber
&&f(z)= \frac{l_A}{\pi} * \frac{r_s}{D_V(z)}*\frac{r_s(z_*)}{r_s(z_d)}=\frac{d_A (z_*)}{D_V(z)},\\
\nonumber
&&\frac{\sigma_{f}}{f}=\sqrt{(\frac{\sigma_{l_A}}{l_A})^{2} + (\frac{\sigma_{BAO}}{r_s(z_*)/D_V(z)})^{2}+(\frac{\sigma_{r_s(z_d)/r_s(z_*)}}{r_s(z_d)/r_s(z_*)})^{2}   }  ,
\end{eqnarray}
where $\sigma$ presents  the error bar.
 And $f$ is regarded to have the ability of  removing the dependence on much of the complex pre-recombination physics which is needed to determine the horizon scale. In the following, we introduce the  BAO and CMB samples separately.

 \begin{table*}[t]
 \tiny
\begin{center}
\begin{tabular}{cccccc}\toprule
Redshift $z$ &$r_{s}(z_d)/D_{V}(z)$ & f(Planck/BAO)  & f(WMAP$9$/BAO) & f(WMAP$9$($w$)/BAO)&  BAO)\\
 \hline $0.106$&$0.336 \pm0.015$ & $31.56\pm1.44$ & $30.82\pm1.43$  &$30.85 \pm1.43$ & Ref. \cite{Beutler:2011hx} 6dFGS\\
 $0.20$& $0.1905\pm 0.0061$ &  $17.95\pm0.60$ & $17.53\pm0.60$ & $17.54 \pm0.60$&Ref.\cite{Percival:2009xn} SDSS LRG\\
  $0.35$& $0.1097  \pm0.0036$  & $10.33\pm0.35$ &$10.09\pm0.35$  & $10.10 \pm0.35$& Ref.\cite{Percival:2009xn} SDSS LRG\\
   $0.275$& $0.1390\pm 0.0037$& $13.09\pm0.37$ &$12.79\pm0.37$ & $12.80 \pm0.37$&Ref.\cite{Percival:2009xn}SDSS LRG\\
   $0.278$& $0.1389\pm 0.0043$  & $13.08\pm0.42$ & $12.78\pm0.42$  &$12.79 \pm0.42$ &Ref.\cite{Kazin:2009cj} SDSS LRG \\
     $0.35$& $0.1126\pm 0.0022$  & $10.61\pm0.23$ &$ 10.36\pm0.24$ &$10.37 \pm0.24$&Ref.\cite{Padmanabhan:2012hf} SDSS LRG\\
   $0.314$& $0.1239\pm 0.0033$  &$11.67\pm0.33$ &$11.40\pm0.33$ & $11.41 \pm0.33$&Ref.\cite{Blake:2011en}   SDSS LRG\\
  $0.44$&$0.0916 \pm0.0071$   & $8.63\pm0.67$ &$8.43\pm0.66$ & $8.44 \pm0.66$ &Ref.\cite{Blake:2011en}  Wigglez\\
  $0.60$&   $0.0726\pm 0.0034$&$6.84\pm0.33$ &$6.68\pm0.32$  &$6.69 \pm0.32$ & Ref.\cite{Blake:2011en}  Wigglez\\
 $0.73$&   $0.0592 \pm0.0032$ &  $5.58\pm0.31$ & $5.45\pm0.30$ & $5.45 \pm0.30$& Ref.\cite{Blake:2011en} Wigglez\\
    $0.32$& $0.1212 \pm0.0024$  & $ 11.42\pm0.25$ & $11.15\pm0.26$& $11.16 \pm0.26$& Ref.\cite{Anderson:2013zyy} BOSS DR11\\
   $0.57$& $0.0732 \pm0.0012$   & $6.90\pm0.13$&$ 6.73\pm0.14$&$6.74 \pm0.14$ & Ref.\cite{Anderson:2013zyy} BOSS DR11\\
   $2.34$& $0.0320 \pm0.0007$  & $ 3.01\pm0.07$ & $2.94\pm0.07$& $2.95 \pm0.07$& Ref.\cite{Delubac:2014aqe}BOSS DR11\\
\hline
\end{tabular}
\end{center}
\caption[crit]{\label{tab1} The BAO and CMB/BAO samples  are derived from 6dFGS, SDSS LRG, WiggleZ, BOSS DR11, Planck and WAMP9. Explanation please see Section \ref{sec2}.}
\end{table*}


\subsection{The BAO distance ratio}
We list the BAO  data $r_s(z_d)/D_V(z)$  in Table \ref{tab1} which  are derived  from the  6dF Galaxy Redshift Survey  (6dFGS), Sloan Digital Sky Survey 
Luminous Red Galaxy sample ((SDSS LRG) , WiggleZ, Baryon Oscillation Spectroscopic Survey (BOSS) DR11 surveys\footnote{ Our BAO data only use the distance ratio,  not the other BAO parameter, e.g. the acoustic parameter $A$. Thus, our BAO-only constraining results is different from the result of Ref.\cite{Addison:2013haa,Aubourg:2014yra,Cheng:2014kja}.}. The $r_s(z_d)/D_V(z)$ data have now been detected over a range of redshifts from $z = 0.106$ to $z = 2.34$.

(1) Beutler et al. analyzed the large-scale correlation function of the 6dFGS and made a 4.5 percent measurement at z = 0.106 \cite{Beutler:2011hx}.

(2) Percival et al. measured the distance ratio
$d_{z} = r_{s}(z_{d})/D_{V }(z)$ at redshifts $z = 0.2$, and $z = 0.35$   by fitting to the power spectra
of luminous red galaxies and main-sample galaxies in the SDSS \cite{Percival:2009xn}. They also showed how the distance-reshift constraints at those two redshifts could be decomposed into a single distance  constant at $z=0.275$ and a  ``gradient''
 around this pivot given by $D_V(0.35)/D_{V}(0.2)$.
 Furthermore,  Kazin et al. examined the correlation function $\xi$ of the SDSS
LRG at large scales using the
final data release and  get the BAO data at $z = 0.278$ \cite{Kazin:2009cj}. Furthermore, Padmanabhan et al. applied the reconstruction technique to the clustering of galaxies from the SDSS  LRG sample, sharpening the BAO feature and achieving the BAO result for  z = 0.35 \cite{Padmanabhan:2012hf}.

(3) Blake et al.  gave  out the  BAO feature in three bins centered at redshifts z = 0.44, 0.60, and 0.73 respectively, using the full sample of $158741$ galaxies from WiggleZ survey.    And they also presented a new measurement of the baryon acoustic feature in the correlation function of the SDSS LRG sample and derived a BAO measurement at  $z=0.314$ that is consistent with previous analyses of the LRG power spectrum \cite{Blake:2011en}.

(4) Fitting for the position of the acoustic features in the correlation function and matter power spectrum of BAO in the clustering of galaxies from
BOSS DR 11, Anderson et al. got   the BAO result for $z=0.32$ and $z=0.57$ \cite{Anderson:2013zyy}.
And from BOSS DR11 latest released sample, Delubac et al. figured out the BAO feature in the flux-correction function of the Lyman-$\alpha$ forest of high reshift quasars    at the effective redshift $z=2.34$ \cite{Delubac:2014aqe}.

\subsection{The acoustic scale}
While the CMB anisotropies have been measured with ever increasing precision by missions such as WMAP$9$ \cite{Bennett:2012zja}, Planck \cite{Ade:2013zuv} and BICEP$2$ \cite{Ade:2014xna}. Sepcially,  the declaration that  the detect
on of the CMB B-mode polarization by the BICEP$2$ collaboration  \cite{Ade:2014xna}   might be wholly or partly due to polarized dust emission \cite{Adam:2014bub}. For conciseness, we choose the Planck and WMAP$9$ data only.
Indeed, the $l_A$ parameter  depends on the background model sightly \cite{Cai:2014ela}.
The used  acoustic scale $l_A(z_{*})$ and $r_s(z_d)/r_s(z_*)$ at the decoupling redshift  derived by  Refs \cite{Wang:2011sb,Wang:2013mha} :
\begin{eqnarray}
\label{cmb}
\nonumber
&&Planck+lensing+WP: l_{A} = 301.57\pm0.18 (0.06\%),  \frac{r_{s}(z_{d})}{r_{s}(z_{*})}=  1.019  \pm    0.009 (0.88\%),\\
\nonumber
&&WMAP9: \,\,\,\, l_{A} = 302.02\pm0.66 (0.22\%),\,\,\, \frac{r_{s}(z_{d})}{r_{s}(z_{*})}=1.045 \pm0.012 (1.15\%),\\
\nonumber
&&WMAP9(w): \,\,\,\, l_{A} = 302.35\pm0.65 (0.21\%),\,\,\, \frac{r_{s}(z_{d})}{r_{s}(z_{*})}=1.045 \pm0.012 (1.15\%).
 \end{eqnarray}
 The First CMB   data combined with Planck lensing, as well as WMAP polarization at low multipoles ($l \leq 23$) \cite{Ade:2013zuv} which represents the tightest constraints from CMB data only at present. 
The second one is derived from the o$\Lambda$CDM model \cite{Wang:2011sb,Wang:2013mha} and WMAP$9$ data. The third one  derived from the  o$w$CDM model \cite{Bennett:2012zja} and WMAP$9$ data.  To distinguish the three kinds of CMB/BAO data, we call them Planck/BAO, WMAP$9$/BAO, and WMAP$9$($w$)/BAO separately.  As Ref.\cite{Cai:2014ela} shows,   the $l_A$ values vary  if changing the fitting model. Our results will show the value of $l_A$ nearly don't affect the constraining results.

 Based on BAO and CMB data, our CMB/BAO sample has 13 data which are presented in Table \ref{tab1}.  The WMAP$9$(w)/BAO data cove the total $5$ CMB/BAO in   Ref.\cite{Li:2013oba}. 
Comparing the error in the BAO data, obviously, the CMB data will bring less uncertainty than BAO to the CMB/BAO, but  the whole CMB/BAO data uncertainty has larger error than the BAO data.  Our results  will show a tighter constraint from the CMB/BAO data compared with BAO data in Table \ref{tab2}.

\subsection{The theoretical models}

 This accelerating behavior of our universe is usually attributed by a presently unknown component, called dark energy, which exhibits negative pressure and dominates over the matter-energy content of our universe. So far, the simplest candidate for dark energy is the cosmological constant. And, the so called  standard cosmological model  $\Lambda$CDM  is in accordance with almost all the existing cosmological observations.
As the  CMB/BAO method retains sensitivity to phenomena that have more effect at higher redshift, such as curvature, we extend the constraining model to $w$CDM  (the  standard cosmological model with constant dark energy equation of state) and o$\Lambda$CDM  (the  standard cosmological model with curvature).   It means we use the following parameters: the equation of state of dark energy w, Hubble parameter H,   and  the dimensionless energy density parameter for the matter $\Omega_{m0}$ and the curvature  $\Omega_{k0}$.

The BAO position measurements  do not provide any $H_0$ constraint, being sensitive only to the combination $H_0 r_s$.  The sound horizon depends on the physical baryon density, $\Omega_b h^{2}$, through $r_s \propto (\Omega_b h^{2})^{-0.13}$.  We then add the prior $\Omega_{b}h^{2} =0.0223\pm 0.0009$  \cite{Pettini:2012ph} to the BAO data  by fixing the CMB mean temperature, which is  fixed to $2.73K$ and determines the energy density in radiation.  On the other side,  the CMB/BAO data is  independent from  the sound horizon, thus CMB/BAO-only data  could not constrain $H_0$.  Correspondingly,  we add the $\Omega_m h^2 = 0.1199\pm 0.0027$ prior from Planck to the CMB/BAO for the $H_0$ constraint.

For data comparison convenience, we divide the data into three samples:(1) the CMB related data Planck/BAO (or +$\Omega_mh^2$), WMAP/BAO (or +$\Omega_mh^2$),  WMAP(w)/BAO (or +$\Omega_mh^2$); (2) the BAO data related data: BAO (or +$\Omega_b h^{2}$),  BAO+Planck/BAO (or +$\Omega_b h^{2}$),  BAO+WMAP/BAO (or +$\Omega_b h^{2}$ ); (3) all combined data: BAO+Planck/BAO +$\Omega_b h^{2}$+$\Omega_m h^{2}$,  BAO+WMAP/BAO +$\Omega_b h^{2}$ +$\Omega_m h^{2}$.
And, we use  the Monte Carlo Markov Chain (MCMC) method based on the publicly available package COSMOMC \cite{Lewis:2002ah} to constrain model parameters which  randomly chooses values for the above parameters, evaluates $\chi^2$
and determines whether to accept or reject the set of parameters by
using the Metropolis-Hastings algorithm.

\begin{table*}[t]
\tiny
\begin{center}
\begin{tabular}{lllll}\toprule
\hline  \multicolumn{5}{c}{The  $\Lambda$CDM Model} \\
 & $\Omega_{m}$ & -&$H_{0}$  & $r_s H_0$ \\
Planck/BAO&  $ 0.256_{-0.010-0.016}^{+ 0.030+ 0.039}$& -&
-&
-
\\
     WMAP$9$/BAO&
     $ 0.277_{-0.015-0.021}^{+ 0.031+ 0.041}$ & -&
-&
-
  \\
   WMAP$9$($w$)/BAO&
   $ 0.277_{-0.015-0.021}^{+ 0.031+ 0.040}$& -&
-&
-
    \\
   
    Planck/BAO+$\Omega_m h^2$&$ 0.271_{-0.011-0.018}^{+ 0.017+ 0.027}$&-&
$ 66.6_{ -2.5 -3.7}^{+  1.8+  2.9}$&-
\\
  WMAP$9$/BAO+$\Omega_m h^2$&$ 0.293_{-0.013-0.020}^{+ 0.018+ 0.029}$& -&
$ 64.0_{ -2.3 -3.3}^{+  1.8+  2.9}$&-
  \\
   WMAP$9$($w$)/BAO+$\Omega_m h^2$&$ 0.292_{-0.013-0.020}^{+ 0.020+ 0.025}$  & 
-&
$ 64.1_{ -2.2 -3.4}^{+  1.8+  2.9}$&
-
    \\
  BAO & 
  $ 0.258_{-0.035-0.054}^{+ 0.039+ 0.066}$ & -&
$ 42.8_{ -4.2 -4.5}^{+ 47.2+ 47.2}$&
$ 0.978_{-0.137-0.150}^{+ 0.216+ 0.233}$

 \\
    BAO+$\Omega_b h^{2}$ &
     $ 0.258_{-0.042-0.061}^{+ 0.049+ 0.076}$& -&
$ 66.2_{ -3.2 -4.6}^{+  3.7+  5.8}$&
$ 1.031_{-0.043-0.063}^{+ 0.046+ 0.071}$

 \\
 BAO+Planck/BAO& 
 $ 0.273_{-0.013-0.020}^{+ 0.016+ 0.025}$& -&
$ 44.9_{ -5.7 -5.7}^{+ 45.1+ 45.1}$&
$ 1.019_{-0.158-0.158}^{+ 0.178+ 0.193}$
 \\
  BAO+Planck/BAO+$\Omega_b h^{2}$&  
  $ 0.275_{-0.015-0.022}^{+ 0.016+ 0.024}$-&
  &$ 67.5_{ -2.5 -3.6}^{+  2.5+  3.9}$&
$ 1.044_{-0.043-0.064}^{+ 0.043+ 0.066}$
\\
  BAO+Planck/BAO+$\Omega_b h^{2}$+$\Omega_m h^{2}$& 
  $ 0.271_{-0.010-0.015}^{+ 0.011+ 0.016}$&-&
$ 66.7_{ -1.4 -2.1}^{+  1.4+  2.1}$&
$ 1.033_{-0.034-0.050}^{+ 0.034+ 0.051}$
  \\
  BAO+WMAP$9$/BAO&
   $ 0.287_{-0.013-0.020}^{+ 0.017+ 0.025}$& -&
$ 45.9_{ -6.5 -6.5}^{+ 44.1+ 44.1}$&
$ 1.027_{-0.159-0.159}^{+ 0.164+ 0.180}$
 \\
 BAO+WMAP$9$/BAO+$\Omega_b h^{2}$& 
 $ 0.290_{-0.015-0.022}^{+ 0.016+ 0.024}$&-&
$ 68.1_{ -2.5 -3.7}^{+  2.6+  4.0}$&
$ 1.042_{-0.044-0.063}^{+ 0.045+ 0.067}$
 \\
  BAO+WMAP$9$/BAO+$\Omega_b h^{2}$+$\Omega_m h^{2}$& 
 $ 0.280_{-0.010-0.015}^{+ 0.011+ 0.016}$&-&
$ 66.0_{ -1.4 -2.0}^{+  1.4+  2.1}$&
$ 1.014_{-0.033-0.049}^{+ 0.034+ 0.050}$
\\
\hline  \multicolumn{5}{c}{The $w$CDM Model} \\
  & $\Omega_{m}$ & $w$&$H_{0}$ & $r_s H_0$\\
\hline
Planck/BAO& 
$ 0.261_{-0.029-0.043}^{+ 0.026+ 0.035}$&
$-1.095_{-0.342-0.541}^{+ 0.295+ 0.419}$&
-&
-
 \\
     WMAP$9$/BAO&
     $ 0.266_{-0.028-0.044}^{+ 0.036+ 0.051}$&
$-1.164_{-0.417-0.659}^{+ 0.292+ 0.426}$&
-&
-
  \\
   WMAP$9$($w$)/BAO&
   $ 0.267_{-0.027-0.044}^{+ 0.037+ 0.051}$&
$-1.139_{-0.435-0.667}^{+ 0.284+ 0.420}$&
-&
-
    \\
        Planck/BAO+$\Omega_m h^2$&
$ 0.263_{-0.030-0.046}^{+ 0.026+ 0.036}$&
$-1.110_{-0.414-0.602}^{+ 0.308+ 0.434}$&
$ 67.5_{ -3.6 -4.6}^{+  4.4+  7.1}$&-
\\
     WMAP$9$/BAO+$\Omega_m h^2$&
$ 0.277_{-0.036-0.054}^{+ 0.033+ 0.047}$&
$-1.232_{-0.440-0.681}^{+ 0.354+ 0.496}$&
$ 65.9_{ -3.9 -5.3}^{+  4.9+  7.9}$&-
  \\
   WMAP$9$($w$)/BAO+$\Omega_m h^2$&
$ 0.277_{-0.036-0.054}^{+ 0.032+ 0.044}$&
$-1.217_{-0.437-0.682}^{+ 0.350+ 0.492}$&
$ 65.8_{ -3.7 -5.2}^{+  4.9+  7.8}$&-
    \\
  BAO &  
  $ 0.262_{-0.058-0.143}^{+ 0.049+ 0.073}$&
$-1.112_{-0.530-0.810}^{+ 0.442+ 0.654}$&
$ 70.6_{-34.5-34.5}^{+ 19.4+ 19.4}$&
$ 1.098_{-0.364-0.413}^{+ 0.322+ 0.443}$
\\
    BAO+$\Omega_b h^{2}$ &  
    $ 0.263_{-0.074-0.192}^{+ 0.055+ 0.081}$&
$-1.122_{-0.609-0.877}^{+ 0.516+ 0.703}$&
$ 69.2_{-15.1-23.3}^{+ 13.9+ 20.5}$&
$ 1.096_{-0.297-0.460}^{+ 0.335+ 0.507}$
    \\
 BAO+Planck/BAO& 
 $ 0.265_{-0.022-0.032}^{+ 0.023+ 0.033}$&
$-1.121_{-0.275-0.408}^{+ 0.224+ 0.317}$&
$ 50.9_{-10.4-11.7}^{+ 39.1+ 39.1}$&
$ 1.096_{-0.183-0.250}^{+ 0.246+ 0.310}$
 \\
  BAO+Planck/BAO+$\Omega_b h^{2}$&
  $ 0.267_{-0.025-0.036}^{+ 0.025+ 0.034}$&
$-1.132_{-0.304-0.441}^{+ 0.255+ 0.349}$&
$ 70.1_{ -5.9 -8.1}^{+  6.7+  9.7}$&
$ 1.111_{-0.141-0.190}^{+ 0.176+ 0.259}$
   \\
     BAO+Planck/BAO+$\Omega_b h^{2}$+$\Omega_m h^{2}$&
     $ 0.267_{-0.024-0.035}^{+ 0.022+ 0.031}$&
$-1.042_{-0.242-0.357}^{+ 0.196+ 0.267}$&
$ 67.3_{ -3.0 -4.2}^{+  3.6+  5.4}$&
$ 1.051_{-0.090-0.120}^{+ 0.116+ 0.173}$
     \\
  BAO+WMAP$9$/BAO&
  $ 0.275_{-0.025-0.036}^{+ 0.024+ 0.035}$&
$-1.219_{-0.312-0.473}^{+ 0.254+ 0.360}$&
$ 47.2_{ -7.4 -7.4}^{+ 42.8+ 42.8}$&
$ 1.123_{-0.236-0.236}^{+ 0.262+ 0.333}$
 \\
 BAO+WMAP$9$/BAO+$\Omega_b h^{2}$& 
 $ 0.276_{-0.028-0.040}^{+ 0.026+ 0.037}$&
$-1.240_{-0.341-0.498}^{+ 0.296+ 0.406}$&
$ 72.7_{ -6.5 -8.9}^{+  7.1+ 10.5}$&
$ 1.163_{-0.156-0.214}^{+ 0.190+ 0.289}$
 \\
  BAO+WMAP$9$/BAO+$\Omega_b h^{2}$+$\Omega_m h^{2}$& 
 $ 0.278_{-0.024-0.035}^{+ 0.022+ 0.031}$&
$-1.009_{-0.256-0.375}^{+ 0.197+ 0.268}$&
$ 66.1_{ -2.8 -3.9}^{+  3.6+  5.3}$&
$ 1.018_{-0.083-0.113}^{+ 0.113+ 0.168}$
\\
 \hline  \multicolumn{5}{c}{The o$\Lambda$CDM Model} \\
  & $\Omega_{m}$ & $\Omega_{k}$&$H_{0}$ & $r_s H_0$ \\
\hline
Planck/BAO&  
$ 0.259_{-0.037-0.057}^{+ 0.038+ 0.066}$&
$-0.010_{-0.020-0.028}^{+ 0.034+ 0.051}$&
-&
-
\\
     WMAP$9$/BAO&
     $ 0.259_{-0.036-0.056}^{+ 0.041+ 0.070}$&
$-0.025_{-0.016-0.024}^{+ 0.036+ 0.054}$
&
-&
-
  \\
   WMAP$9$($w$)/BAO&
   $ 0.262_{-0.038-0.057}^{+ 0.041+ 0.069}$&
$-0.022_{-0.015-0.025}^{+ 0.035+ 0.052}$
&
-&
-
    \\
        Planck/BAO+$\Omega_m h^2$& $ 0.257_{-0.041-0.059}^{+ 0.049+ 0.077}$&
$-0.009_{-0.024-0.032}^{+ 0.031+ 0.049}$&
$ 68.3_{ -5.9 -8.6}^{+  6.4+  9.9}$&-
\\
     WMAP$9$/BAO+$\Omega_m h^2$&
$ 0.259_{-0.043-0.061}^{+ 0.050+ 0.079}$&
$-0.021_{-0.022-0.029}^{+ 0.030+ 0.047}$&
$ 68.0_{ -5.9 -8.6}^{+  6.6+ 10.2}$&-
  \\
   WMAP$9$($w$)/BAO+$\Omega_m h^2$&
$ 0.262_{-0.044-0.062}^{+ 0.051+ 0.079}$&
$-0.018_{-0.023-0.031}^{+ 0.029+ 0.047}$&
$ 67.7_{ -5.9 -8.6}^{+  6.6+ 10.1}$&-
    \\ 
  BAO &  
  $ 0.268_{-0.058-0.094}^{+ 0.051+ 0.077}$&
$-0.103_{-0.256-0.343}^{+ 0.391+ 0.653}$
&
$ 76.0_{-38.9-38.9}^{+ 14.0+ 14.0}$&
$ 1.131_{-0.358-0.365}^{+ 0.247+ 0.320}$
\\
    BAO+$\Omega_b h^{2}$ & 
    $ 0.266_{-0.064-0.096}^{+ 0.059+ 0.085}$&
$-0.097_{-0.285-0.368}^{+ 0.438+ 0.688}$
&
$ 69.2_{-12.4-17.4}^{+ 12.9+ 18.1}$&
$ 1.096_{-0.252-0.349}^{+ 0.259+ 0.360}$
 \\
 BAO+Planck/BAO& 
 $ 0.259_{-0.030-0.043}^{+ 0.033+ 0.051}$&
$-0.021_{-0.016-0.023}^{+ 0.021+ 0.031}$
&
$ 86.3_{-45.2-46.9}^{+  3.7+  3.7}$&
$ 1.154_{-0.222-0.281}^{+ 0.056+ 0.072}$
 \\
  BAO+Planck/BAO+$\Omega_b h^{2}$& 
  $ 0.258_{-0.033-0.045}^{+ 0.040+ 0.058}$&
$-0.021_{-0.018-0.024}^{+ 0.023+ 0.034}$
&
$ 66.6_{ -3.2 -4.5}^{+  3.7+  5.3}$&
$ 1.042_{-0.047-0.067}^{+ 0.052+ 0.074}$
  \\
    BAO+Planck/BAO+$\Omega_b h^{2}$& 
     $ 0.265_{-0.016-0.022}^{+ 0.017+ 0.024}$&
$-0.017_{-0.013-0.018}^{+ 0.014+ 0.020}$&
$ 67.2_{ -1.9 -2.6}^{+  1.9+  2.7}$&
$ 1.047_{-0.046-0.065}^{+ 0.046+ 0.066}$
    \\
  BAO+WMAP$9$/BAO&
  $ 0.260_{-0.030-0.043}^{+ 0.033+ 0.051}$&
$-0.021_{-0.016-0.022}^{+ 0.021+ 0.032}$
&
$ 86.0_{-44.9-46.4}^{+  4.0+  4.0}$&
$ 1.151_{-0.214-0.277}^{+ 0.059+ 0.073}$
 \\
 BAO+WMAP$9$/BAO+$\Omega_b h^{2}$& 
 $ 0.259_{-0.034-0.046}^{+ 0.039+ 0.057}$&
$-0.020_{-0.018-0.024}^{+ 0.023+ 0.033}$&
$ 66.7_{ -3.3 -4.6}^{+  3.6+  5.3}$&
$ 1.044_{-0.049-0.069}^{+ 0.050+ 0.073}$
 \\
  BAO+WMAP$9$/BAO+$\Omega_b h^{2}$+$\Omega_mh^2$& 
  $ 0.265_{-0.015-0.022}^{+ 0.017+ 0.024}$&
$-0.018_{-0.012-0.018}^{+ 0.014+ 0.020}$&
$ 67.3_{ -1.9 -2.7}^{+  1.8+  2.6}$&
$ 1.048_{-0.047-0.066}^{+ 0.046+ 0.065}$

\\
\hline
\end{tabular}
\end{center}
\caption[crit]{\small The  constraining results for the $\Lambda$CDM, o$\Lambda$CDM and $w$CDM models.    \label{tab2}}
\end{table*}

\begin{figure}
  \begin{center}
   {\includegraphics[width=2.0in]{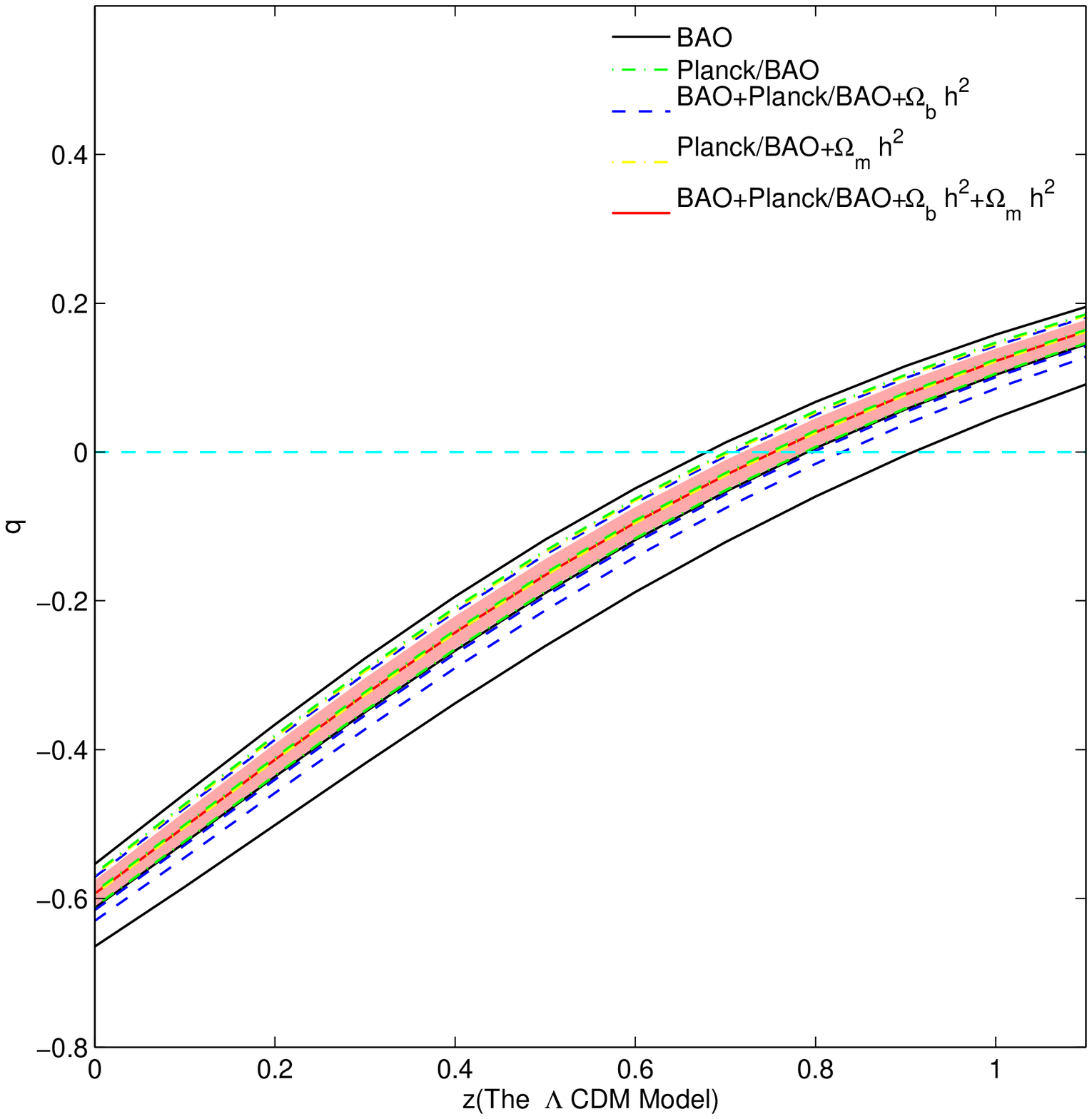}}\quad
        {\includegraphics[width=2.0in]{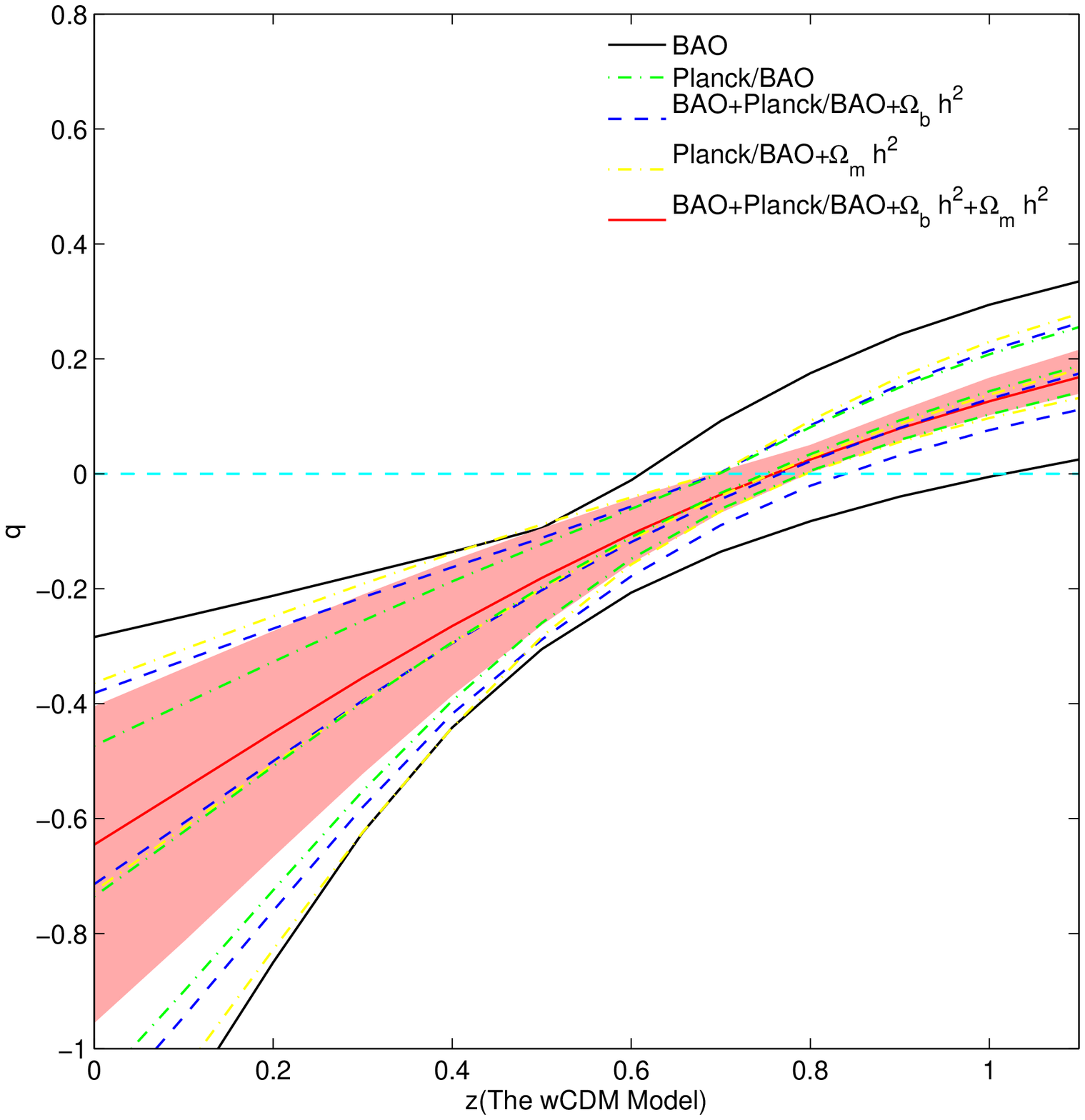}}\quad
             {\includegraphics[width=2.0in]{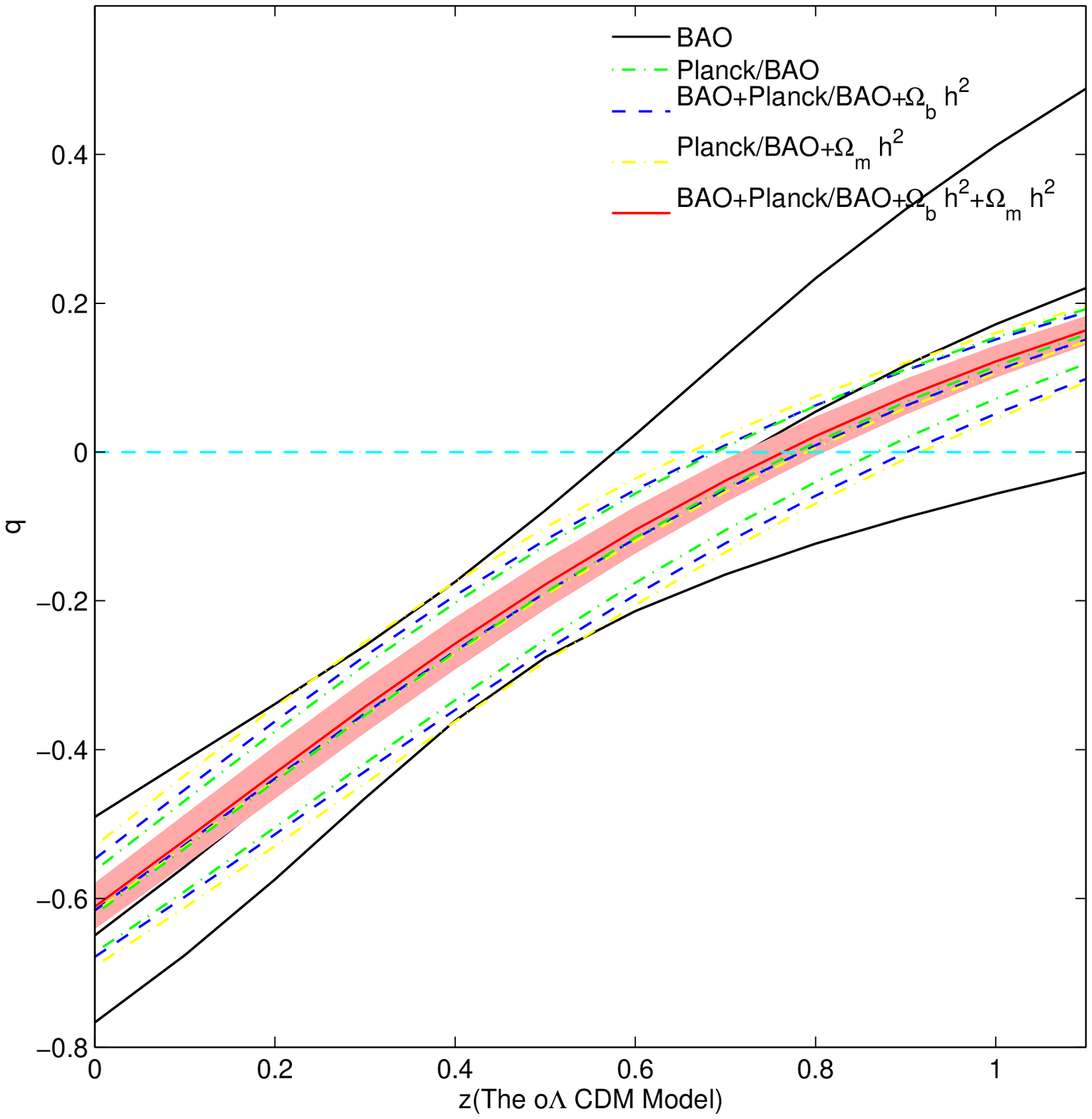}}\quad
    \end{center}
 \caption{\small The evolution of the deceleration parameter q vs the redshift z for selected data.  }\label{q}
\end{figure}

\section{Data Discussions and Parameter Comparisons}\label{sec3}
 We   report mean parameter values and  boundaries of the symmetric 68$\%$ and 95$\%$ (1$\sigma$ and 2$\sigma$ confidence intervals (C.L.))  for all the  models in Table \ref{tab2}.  
 Before further analysis, we explain the value of $\chi^2$ firstly.  The BAO-only data obtain $\chi^{2}=2.272$ in the  $\Lambda$CDM model, and  the Planck/BAO data give $\chi^2=1.986$. Considering the numbers of the  BAO and CMB/BAO data  are both 13, the $\chi^2$ is too small.  And,  limited  improvement of $\chi^{2}$ is given out  after adding the $\Omega_b h^2$ prior and extending the  $\Lambda$CDM model. 
 Ref.\cite{Addison:2013haa} concluded this phenomenon is because parricidal overlap in both redshift and sky coverage for WiggleZ and BOSS. As we neglect  interdependence between constraints from different surveys,
our overlap dataset also have the same small $\chi^{2}$.

Basically, all the data are effective to  give out an accelerating universe which are shown in  Figure \ref{q}.   The derived  
 decelerating parameter 
$q(z)=- a\ddot{a}/\dot{a}^{2}$ represents an accelerating universe when  $-1<q_{0}<0$. And the transition redshift  $z_t$  where $q(z_t)=0 $  denotes the time when our universe evolved from cosmic deceleration ($q > 0$) to acceleration($q < 0$).
  As a representative, the Planck/BAO data obtain
  \begin{eqnarray}
  \nonumber
 &&  q_{0}=-0.62_{ -0.01 -0.02}^{+  0.04+  0.06 }, z_t=0.80_{ -0.09 -0.11}^{+  0.03+  0.05}  ( \Lambda CDM),\\
 \nonumber
 &&q_0=-0.71_{ -0.41 -0.68}^{+  0.33+  0.47 },  z_t= 0.77_{ -0.07 -0.13}^{+  0.07+  0.09} (wCDM),\\
 \nonumber
  &&   q_0=-0.62_{ -0.06 -0.10}^{+  0.07+  0.12 }, z_t=0.80_{ -0.13 -0.21}^{+  0.13+  0.22}  (o \Lambda CDM).
     \end{eqnarray}
  $q_0$ and $z_t$ constraints have nearly identical central value to the most constraining result of $BAO+Planck/BAO+\Omega_m h^2+/\Omega_b h^2$, but with larger error. 
   Specially, the $w$CDM model gives out the latest transfer redshift and its deceleration parameter is smaller than the ones in the $\Lambda$CDM and o$\Lambda$CDM models. It is because the fitting value of $w$ is less than $-1$, the universe is accelerating faster.
   
    In general, the  $\Lambda$CDM give out the tightest constraint while the extended parameter w and $\Omega_{k0}$ enlarge the  parameter space.  Although the chosen procedure is different, it is clear that  the BAO+CMB/BAO+$\Omega_b h^2$+$\Omega_m h^2$ data make the main effect of tightening the parameter region. And to   investigate  the CMB and  BAO effect separately, we plot the best fit value and  parameter contours in Figure \ref{m}.    The constraining results for all the models are nearly the same between WMAP$9$/BAO and WMAP$9$($w$)/BAO data for $\Omega_{m0}$, and just  have a small difference for    $\Omega_{k0}$ and $w$. The WMAP$9$($w$)/BAO data could be replaced by the WMAP$9$/BAO data.  Meanwhile, there are obvious shifts between the  Planck/BAO and WMAP$9$/BAO data for the  $\Omega_{m0}$, $\Omega_{k0}$ and $w$ parameters.
   The Planck  and WMAP$9$ survey bring different $l_A$ and $r_s(z_d)/r_s(z_*)$ while  the WMAP$9$/BAO and WMAP$9$($w$)/BAO have the same  $r_s(z_d)/r_s(z_*)$.  The result of  Planck/BAO, WMAP$9$/BAO and WMAP$9$($w$)/BAO  are  expected because the error  for      $r_s(z_d)/r_s(z_*)$   is $1\%$ while that for $l_A$ is $0.02\%$.
   And,
the different results between Planck/BAO and WMAP$9$/BAO show through the recombination history  not related to  the CMB/BAO data directly, it determined the value and error of the CMB/BAO data. 

\begin{figure}
  \begin{center}
        {\includegraphics[width=1.9in]{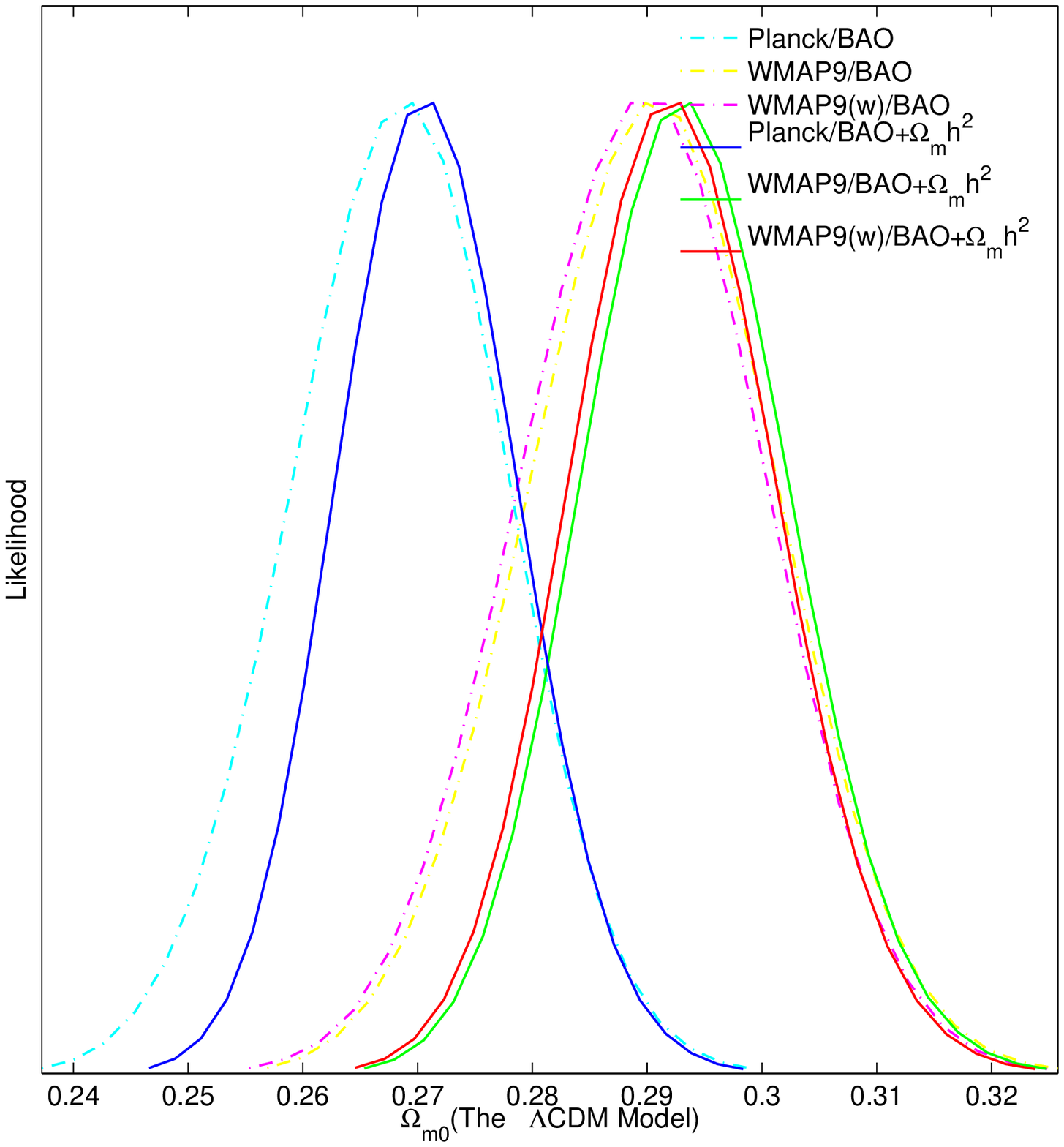}}\quad
         {\includegraphics[width=1.9in]{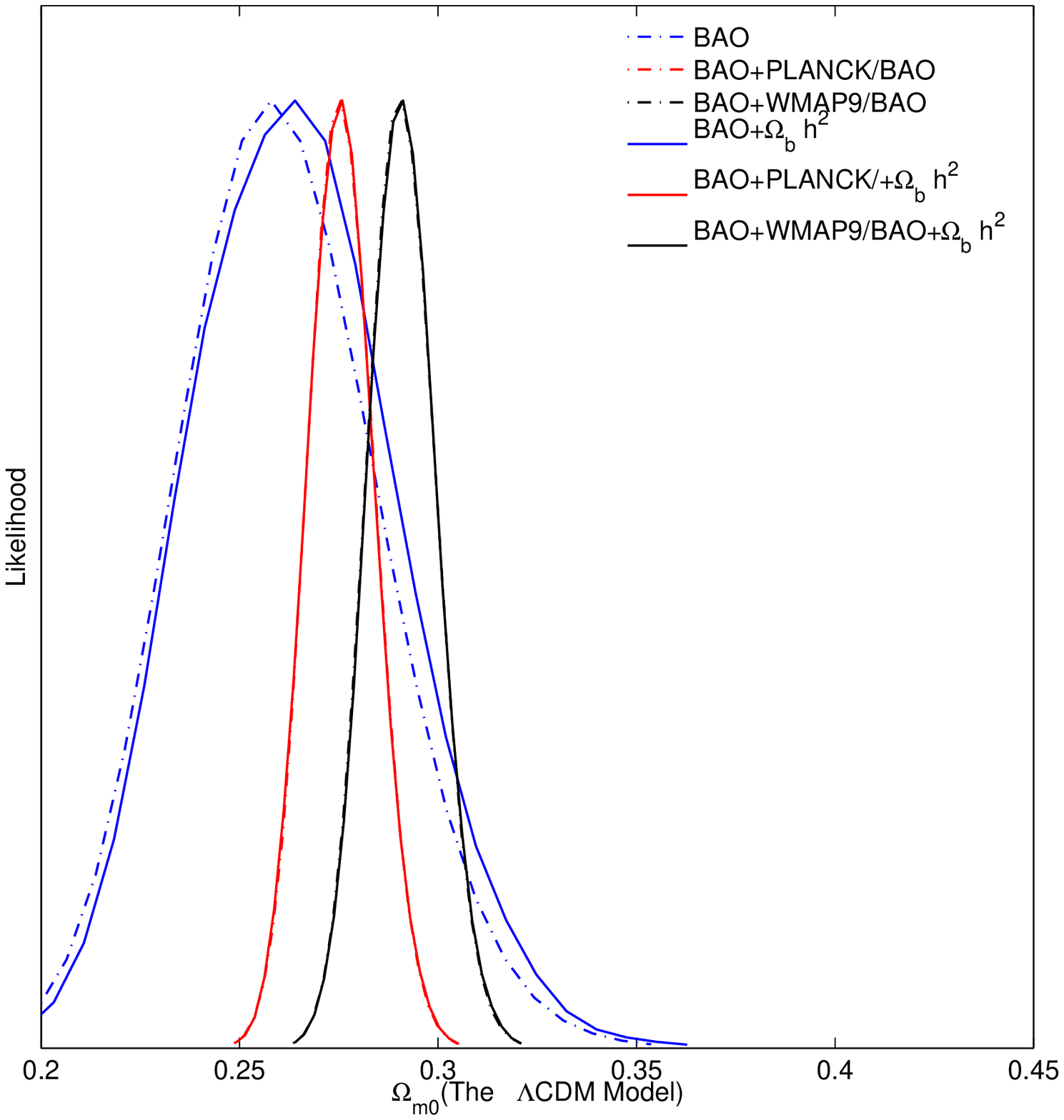}}\quad
           {\includegraphics[width=1.9in]{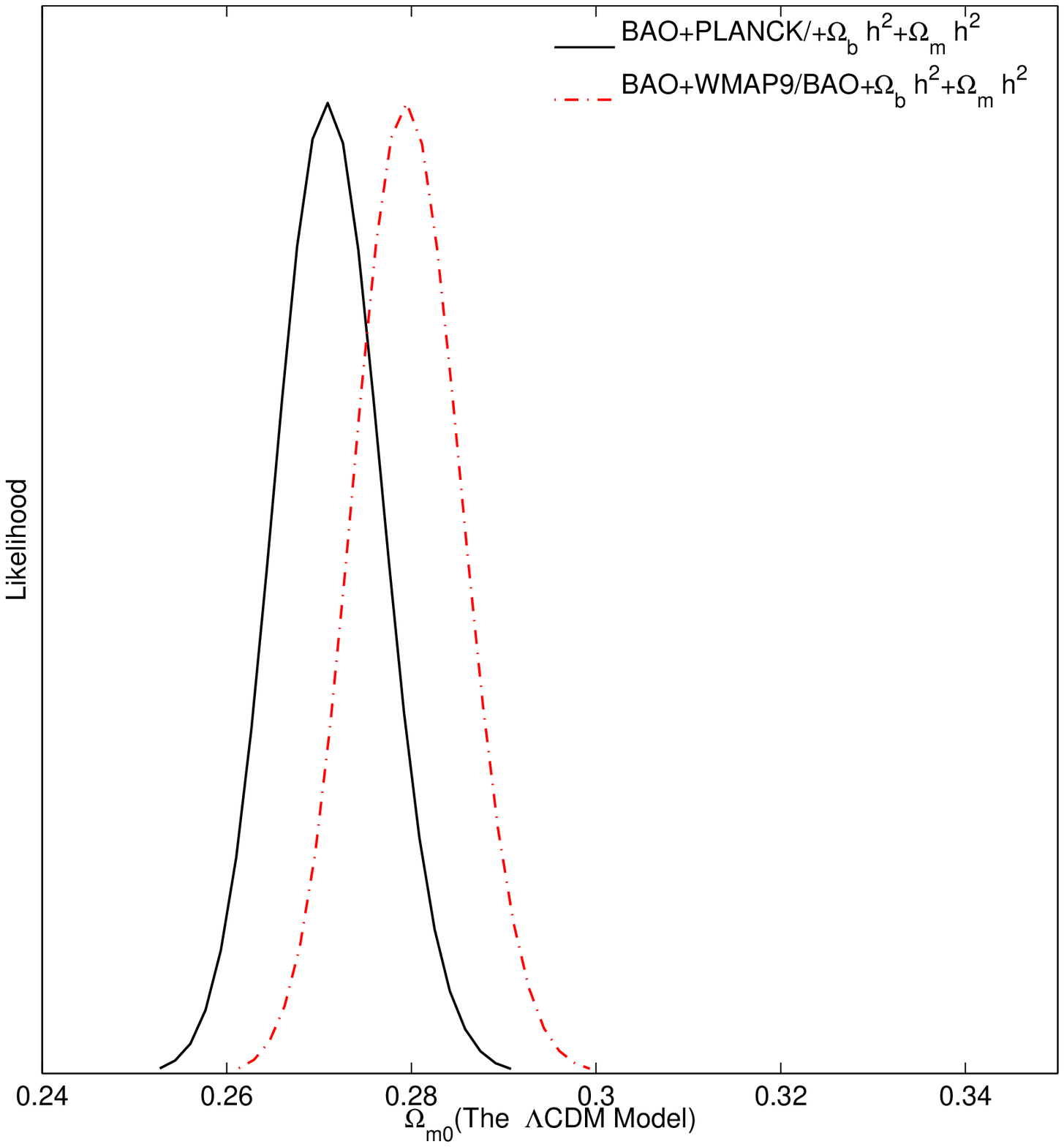}}\quad
             {\includegraphics[width=2.0in]{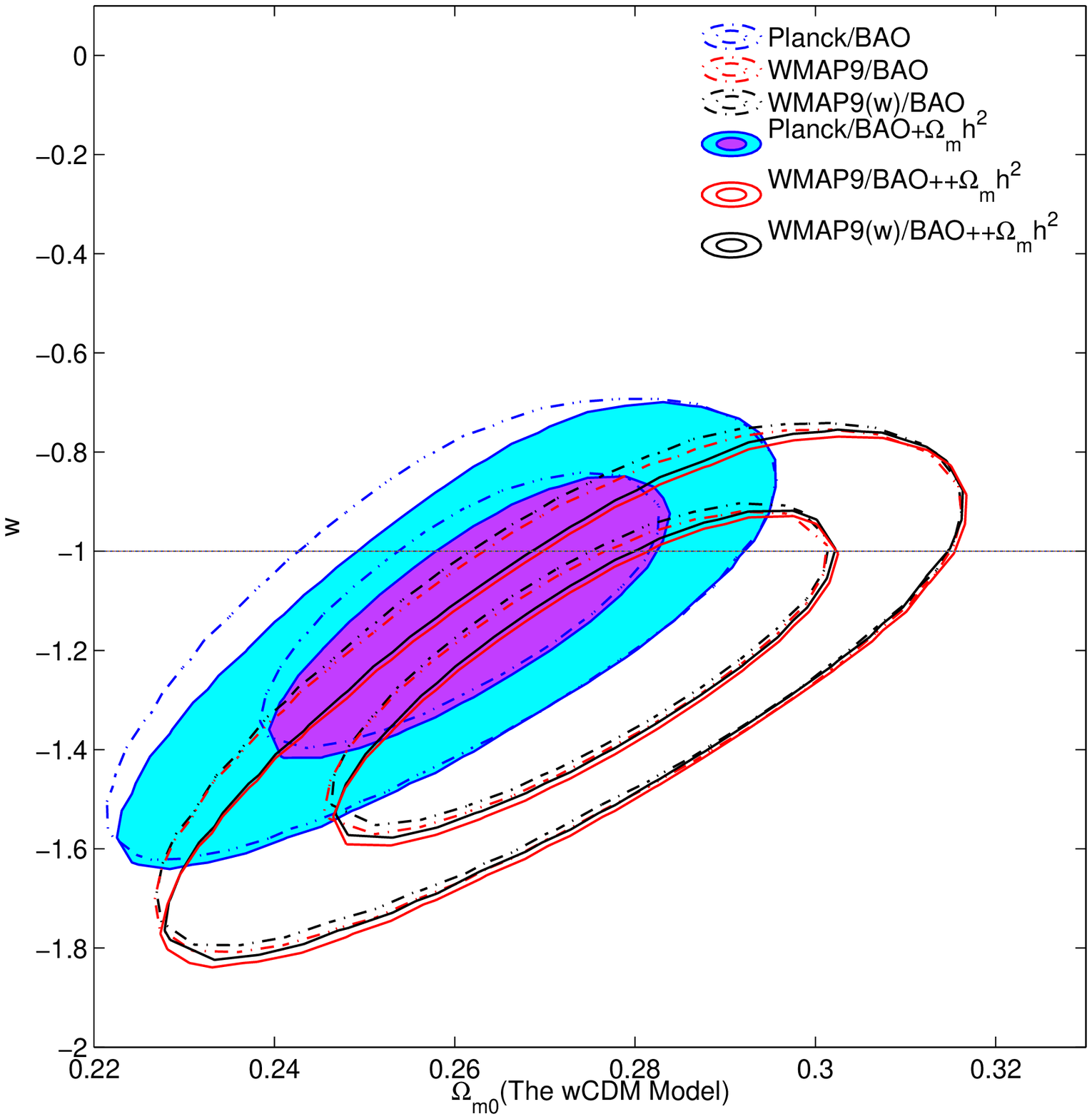}}\quad
                    {\includegraphics[width=2.0in]{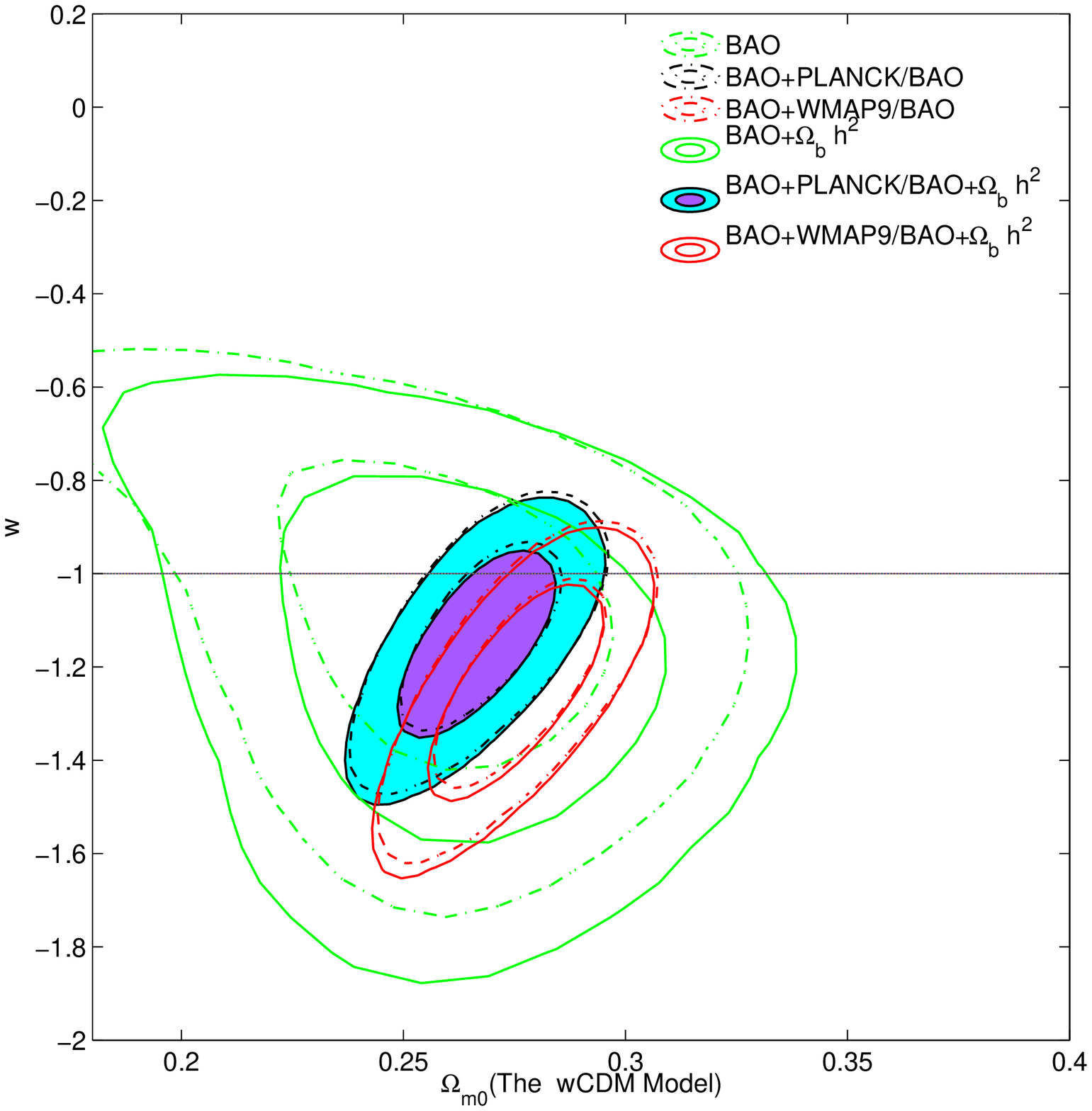}}\quad
                      {\includegraphics[width=2.0in]{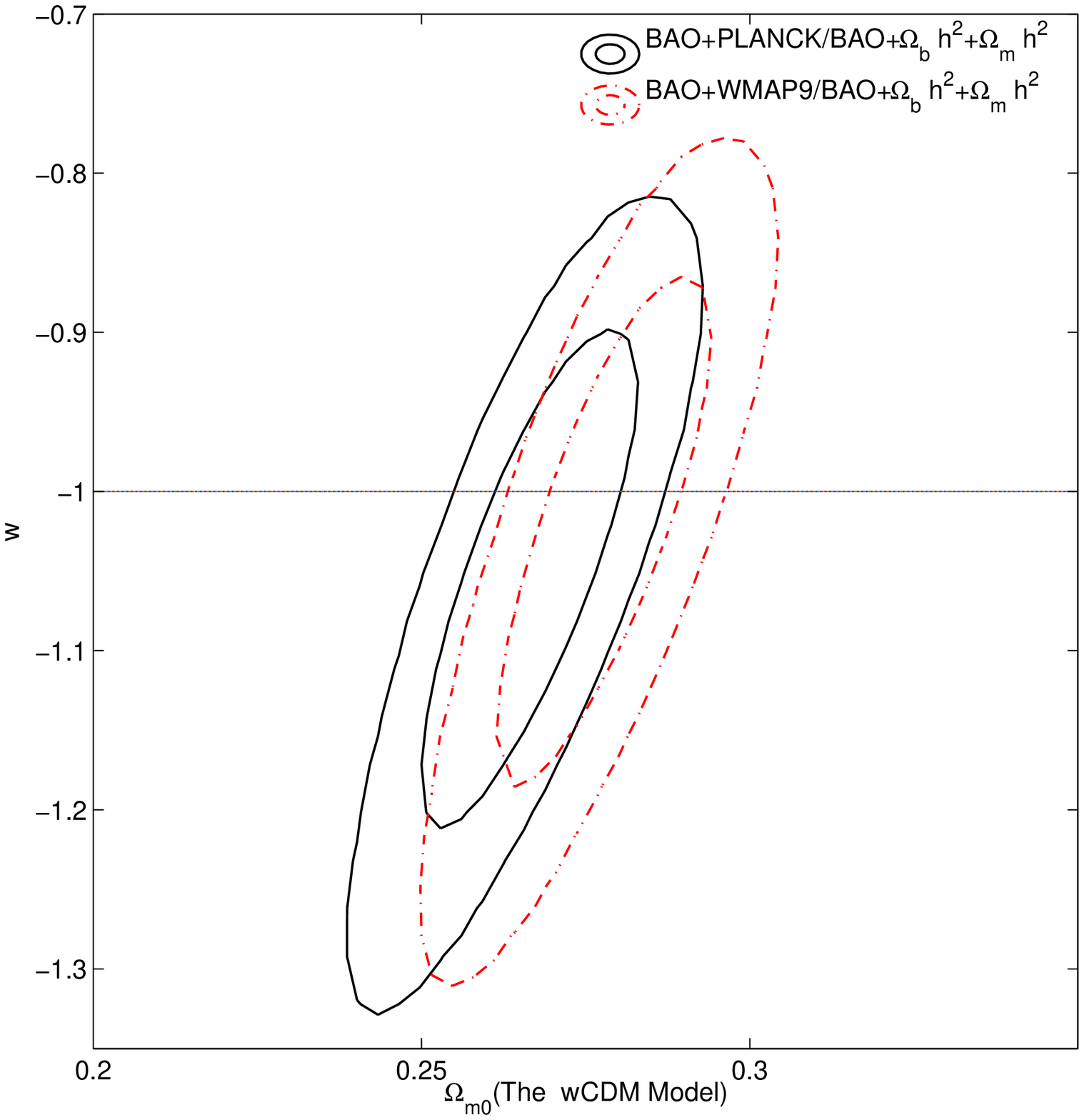}}\quad
                  {\includegraphics[width=2.0in]{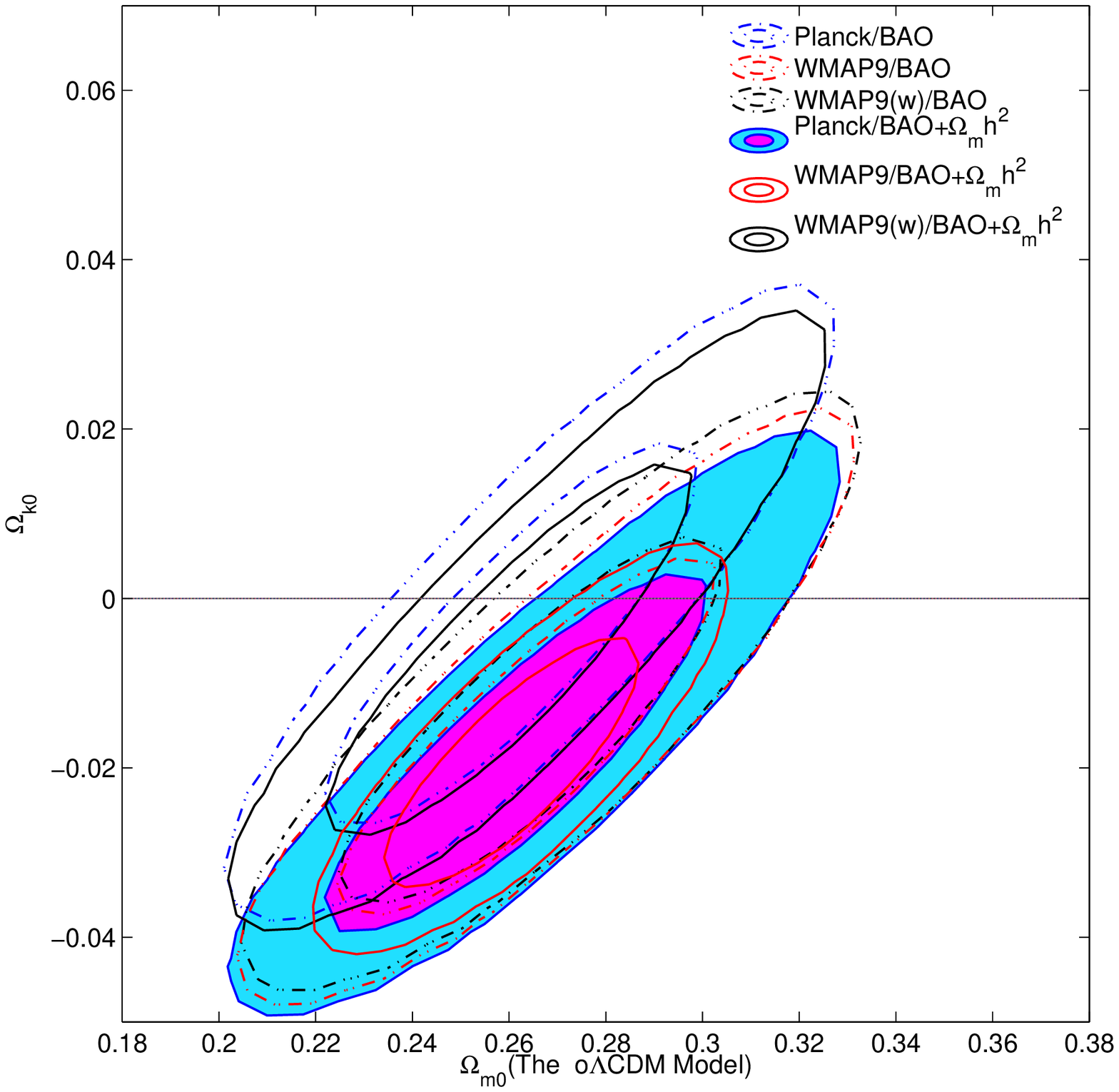}}\quad
                 {\includegraphics[width=2.0in]{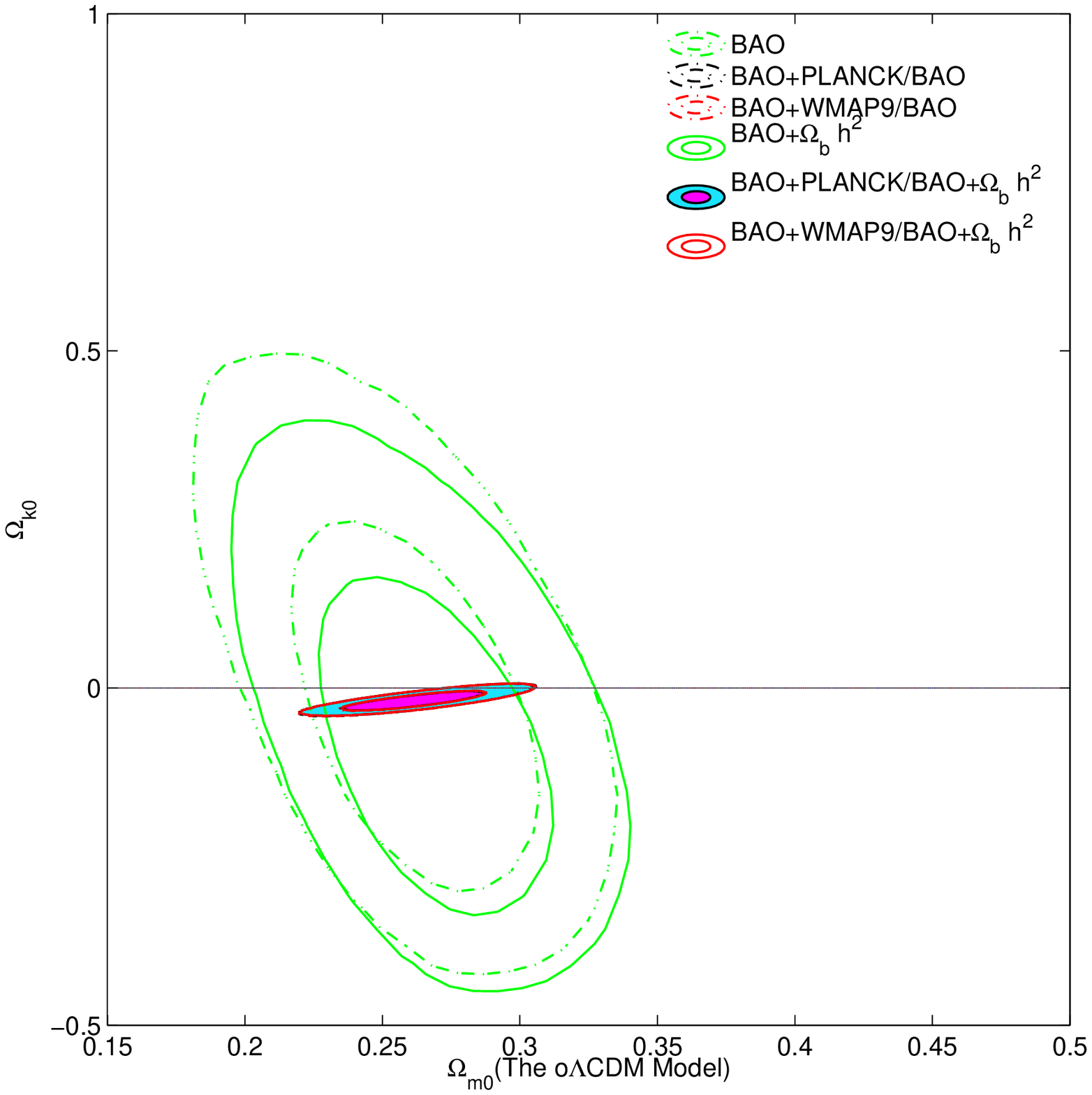}}\quad
                   {\includegraphics[width=2.0in]{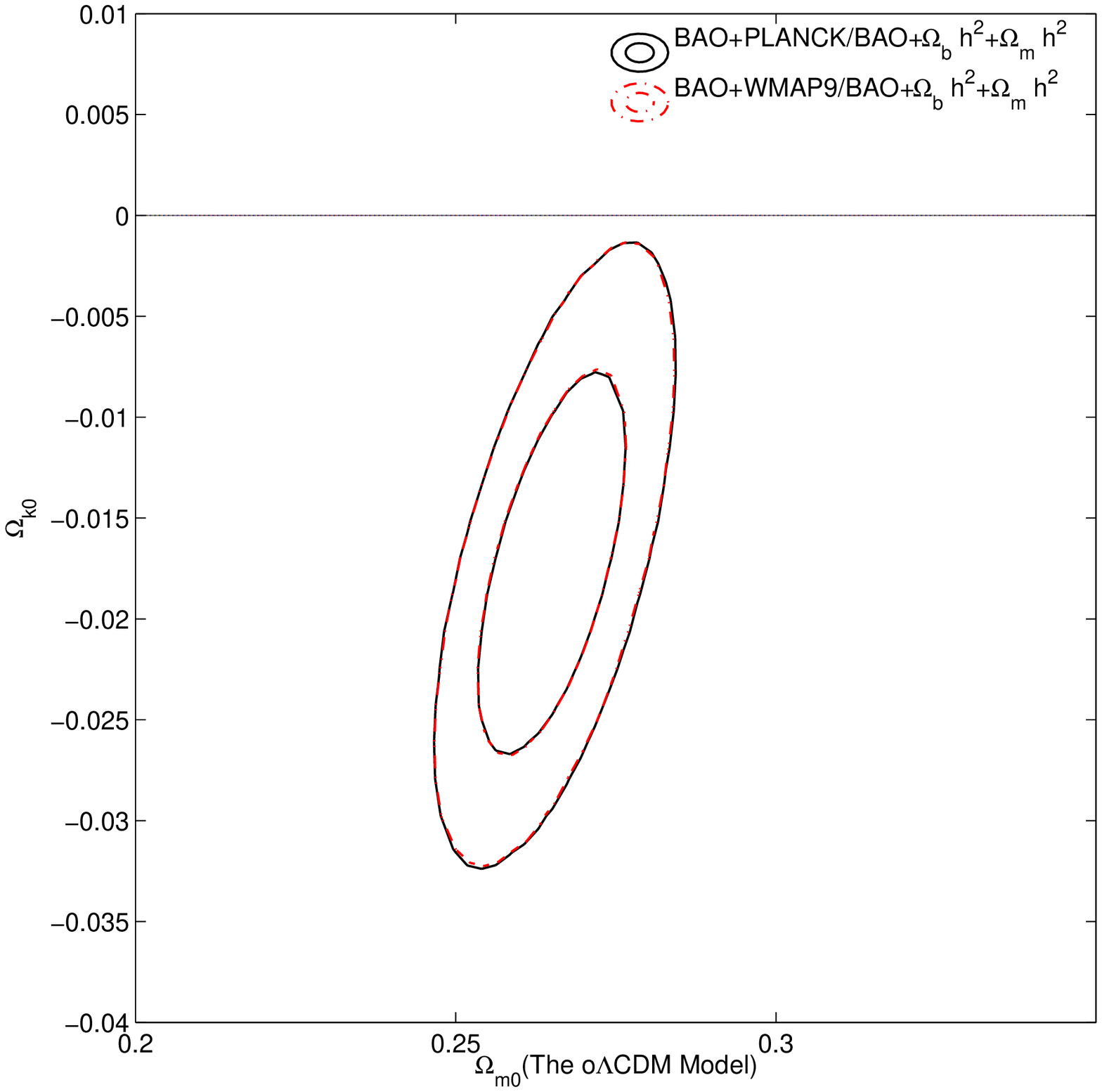}}\quad
    \end{center}
 \caption{\small The upper three panels are  the  values of the likelihood of   the parameter $\Omega_m$ for the  $\Lambda$CDM model. The middle three panels in the transverse direction are the  contour plots  of $\Omega_{m0}-w$ for the $w$CDM model. The lower three panels are the   contour plots of $\Omega_{m0}-\Omega_{k0}$ for the o$\Lambda$CDM model.  The left, middle and right three panels are for CMB comparison, BAO comparison and tightest constraint displaytion.  The lines $w=-1$ and $\Omega_{k0}=0$ show the fixed values in the $\Lambda$CDM model. }\label{m}
\end{figure}

\subsection{The $\Omega_{m0}$ parameter}
For the  $\Lambda$CDM model, the BAO+Planck/BAO+$\Omega_{b}h^2$+$\Omega_{m}h^2$ data give out $\Omega_{m0}=  0.271_{-0.010-0.015}^{+ 0.011+ 0.016}$ which has a obvious tension with the  Planck+WP+highL+BAO result where
$\Omega_{m0}=0.308\pm0.010 (68\%)$.
And, our result  has a tension with the SNLS data  which shows  $\Omega_{m0}= 0.227_{-0.035}^{+0.042} (68\%)$, so is the Planck result.  The tension between Planck and SNLS could be regarded  as the systematics in SNLS SNe IA data, so it could be also used to explain  pour results. And our results is consistent with  the  Union2.1 data where $\Omega_{m0}=0.295^{+0.043}_{-0.040} (68\%)$  and the JLA data where $\Omega_{m0}=0.295\pm0.034 (68\%)$\footnote{ JLA is obtained from the joint analysis of the SDSS-II and SNLS ( Supernova Legacy Survey three year sample)  collaborations.}.
The tension of $\Omega_{m0}$ between the Planck data and our  results could be alleviated by extending parameters. 
The  BAO+Planck/BAO+$\Omega_{b}h^2$+$\Omega_{m}h^2$ data give out $\Omega_{m0}= 0.267_{-0.024-0.035}^{+ 0.022+ 0.031}$ for the  $w$CDM model.
Anyway, the final solution to the tension problem should consider  the recombination history because the best fit of $\Omega_{m0}$  of  the BAO+WMAP9/BAO+$\Omega_{b}h^2$+$\Omega_{m}h^2$ data is shifted  for the $\Lambda$CDM and $w$CDM models compared to the BAO+Planck/BAO+$\Omega_{b}h^2$+$\Omega_{m}h^2$ data.

\subsection{The extended parameter $w$ and  $\Omega_{k0}$}

 Both the CMB/BAO-only  and BAO-only data  give a very weak constrain on $\Omega_{k0}$ and $w$.  Luckily, they yield different degeneracy directions for   $\Omega_{m0}-w$ and $\Omega_{m0}-\Omega_{k0}$ which
 are slightly positive degenerated in CMB/BAO which means when $\Omega_{m0}$ inceases, $w$ increases, but negative degenerated  in BAO means when $\Omega_{m0}$ increases, $w$  decreases.     The BAO+CMB/BAO data give out a much tighter constraints and favor the CMB/BAO direction  slightly for both $\Omega_{k0}$ and $w$  as Figure \ref{m} shows.

 And due to the two-dimensional geometric degeneracy, the Planck+WP+BAO data alone constrain  the  range of  the EOS of dark energy  as  $w=-1.13_{-0.25}^{+0.24} (95\%)$. Similarliy,  our results of BAO+Planck/BAO+$\Omega_bh^2$+$\Omega_mh^2$ show $w=-1.042_{-0.242-0.357}^{+0.196+0.267}$.  Comparing the contours  and the $w=-1$ line in Figure \ref{m}, our results slightly favor  phantom where $w<-1$.

The CMB  curvature power spectrum measurements suffer from a well-known ``geometrical degeneracy" which  is broken  via the integrated Sachs-Wolfe (ISW) effect on large angular scales and gravitational lensing of the CMB spectrum. And with the addition of probes of late time physics,    the geometrical degeneracy can be broken   as well. 
Our CMB/BAO results show it  could constrain the curvature effectively which favor a small negative $\Omega_{k0}$  with the precision of $10^{-2}$.
 The accuracy of the BAO+Planck/BAO+$\Omega_bh^2$+$\Omega_mh^2$ results which show $\Omega_{k0}= -0.017_{-0.018}^{+ 0.020} (95\%)$ is close to the Planck+Lensing+WP+highL result where $ \Omega_{k0}=-0.01^{+0.018}_{-0.019} (95\% )$, but has larger error than the Planck+Lensing+WP+highL+BAO results where  $  \Omega_{k0}=-0.001^{+0.0062}_{-0.0065} (95\% )$.

\subsection{The $H_0$ parameter}
 Adding the $\Omega_bh^2$ (or $\Omega_mh^2$) prior only affects the $\Omega_{m0}$  $w$ and $\Omega_{k0}$ parameters slightly as Figures \ref{m} shows, but it affects the $H_0$ parameter heavily  as Figure  \ref{h} shows.
To do  effective containing, we set a range of $H_0$: $30\,km\, s^{-1} \,Mpc^{-1} \leq H_0 \leq 90km s^{-1} Mpc^{-1}$.
Table \ref{tab2} shows the BAO data only give a lower bound to the $H_0$ which is around $40$.   After plus the $\Omega_b h^2$ prior,   the accuracy of $H_0$ is increased  to $5\%$ as Figures \ref{h}  shows. And, the constraint tendency between  $H_0r_s$ and $H_0$  are the same which indicates the $H_0$ constraint is brought by the $\Omega_bh^2$ prior.  
On the other hand, the CMB/BAO data could not constrain the $H_0$ before adding the   $\Omega_m h^2$ prior.  The Planck/BAO+$\Omega_m h^2$ gives out the $ 66.6_{ -2.5 -3.7}^{+  1.8+  2.9} \,km\, s^{-1} \,Mpc^{-1}$ which is tighter than the BAO+  $\Omega_b h^2$ data. 
 As for the degeneracy between $\Omega_{m0}-H_0$, Figure  \ref{h}
 shows  BAO and CMB/BAO related data yield different degeneracy directions.
They are slightly positive corrected in BAO related data, but negative related in CMB/BAO.   
 And the CMB/BAO+BAO  data favor the CMB/BAO direction.

The  Planck+WP+highL results get $H_0= (67.3\pm 1.2) \,km\, s^{-1} \,Mpc^{-1}$ (68\%).  And,  the local distance ladder measurements  obtain $H_0= (73.8\pm 2.4) km s^{-1} Mpc^{-1}$ measured using Cepheid variable stars and low-redshift type IA SNe observed with the Hubble Space Telescope (HST) by Refs. \cite{Riess:2011yx,Freedman:2012ny}.  
Our  $BAO+Planck/BAO +\Omega_b h^2+\Omega_m h^2$ results show $ 66.7_{ -1.4 -2.1}^{+  1.4+  2.1}\,km\, s^{-1} \,Mpc^{-1}$ which does not have tension with the Planck results, but has mild tensions with local distance ladder measurement of $H_0$ in the context of  $\Lambda$CDM model.
  The tension come either from some sources of unknown systematic errors in some astrophysical measurements or the wrong $\Lambda$CDM model applied in fitting the data. After adding the parameter $\Omega_{k0}$ and $w$, this tension is alleviated slightly in our results where the parameter $\Omega_{k0}$ and $w$ enlarge the $H_0$ range.

\begin{figure}
  \begin{center}
        {\includegraphics[width=2.0in]{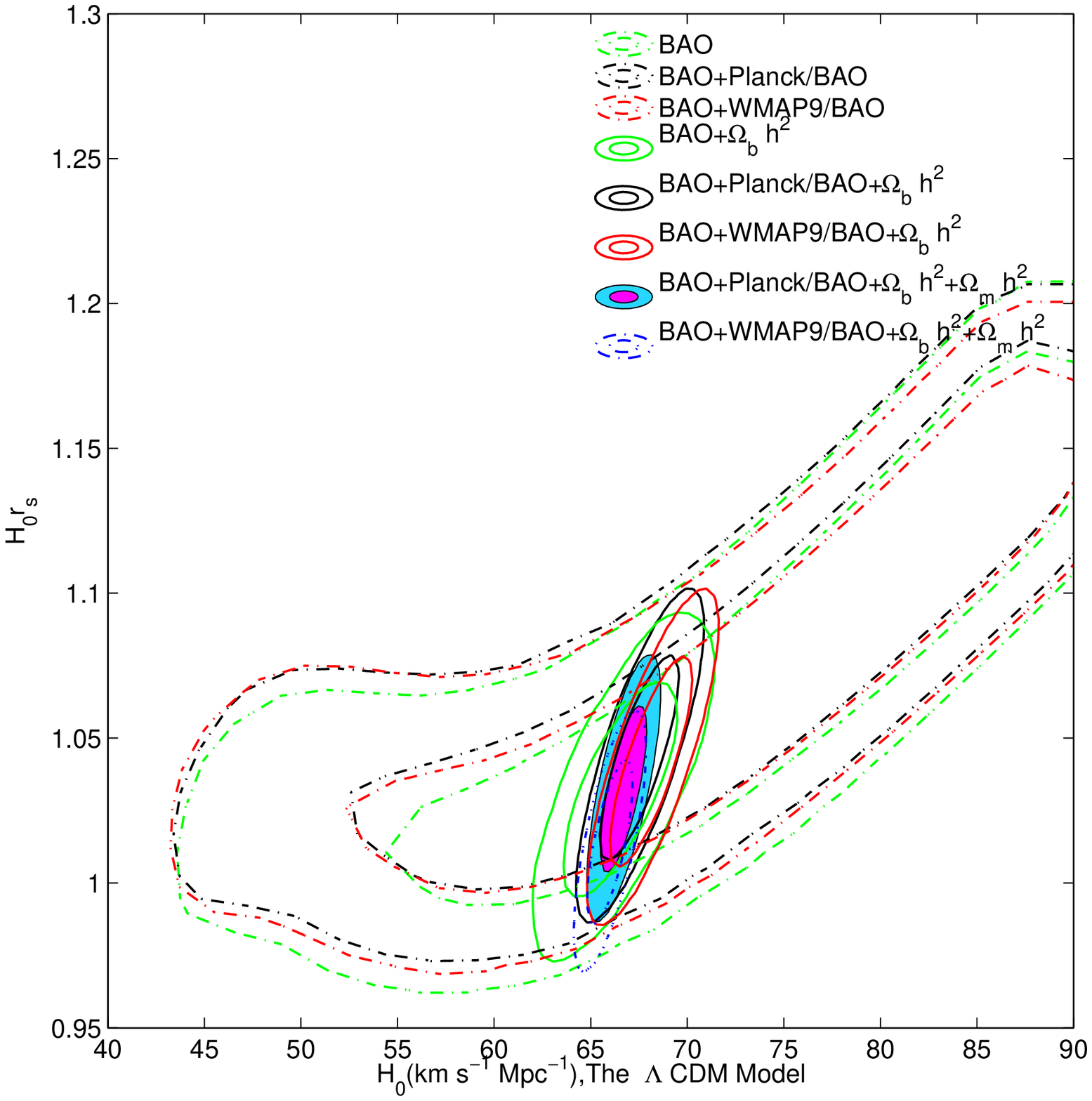}}\quad
             {\includegraphics[width=2.0in]{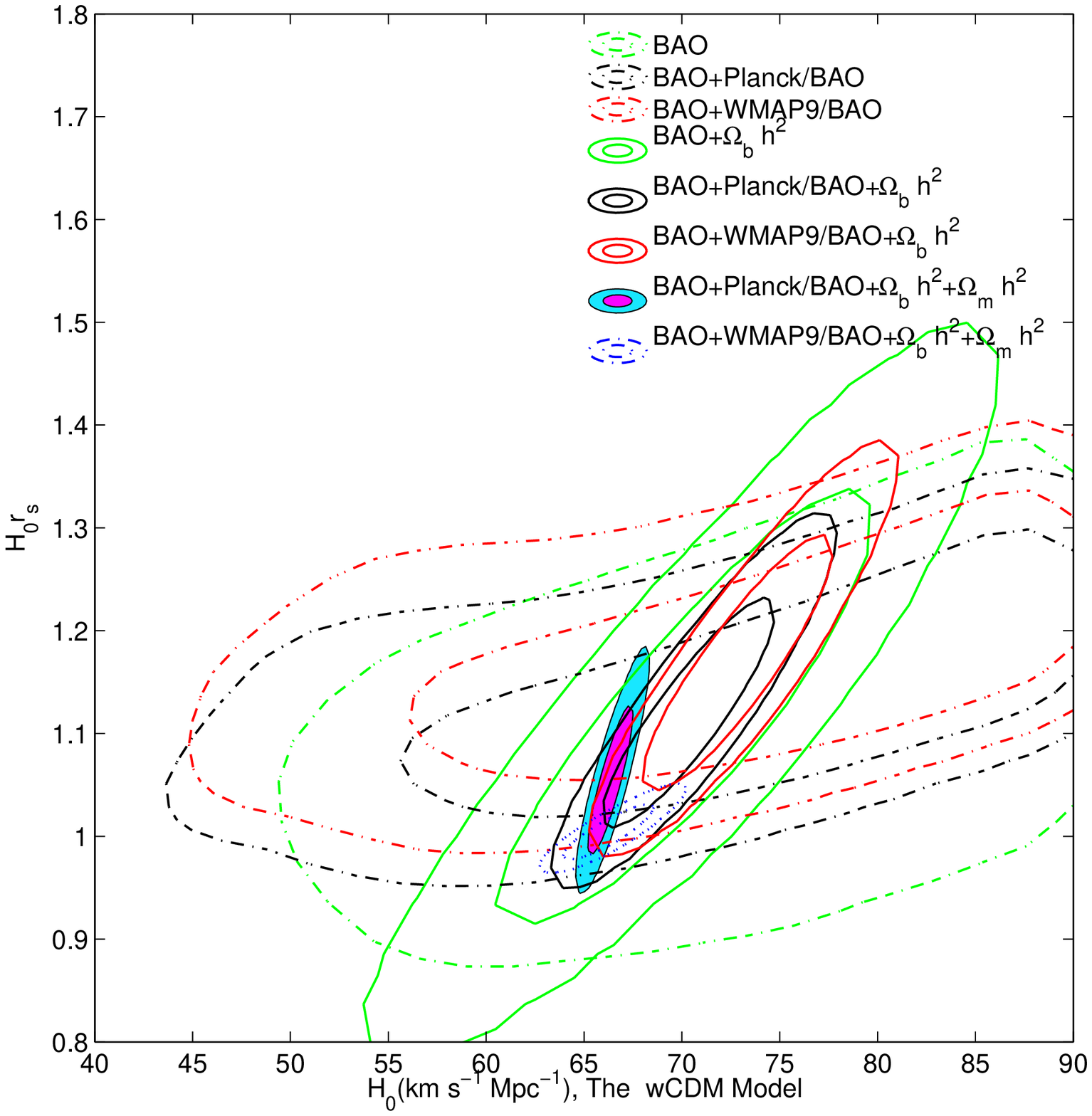}}\quad
                  {\includegraphics[width=2.0in]{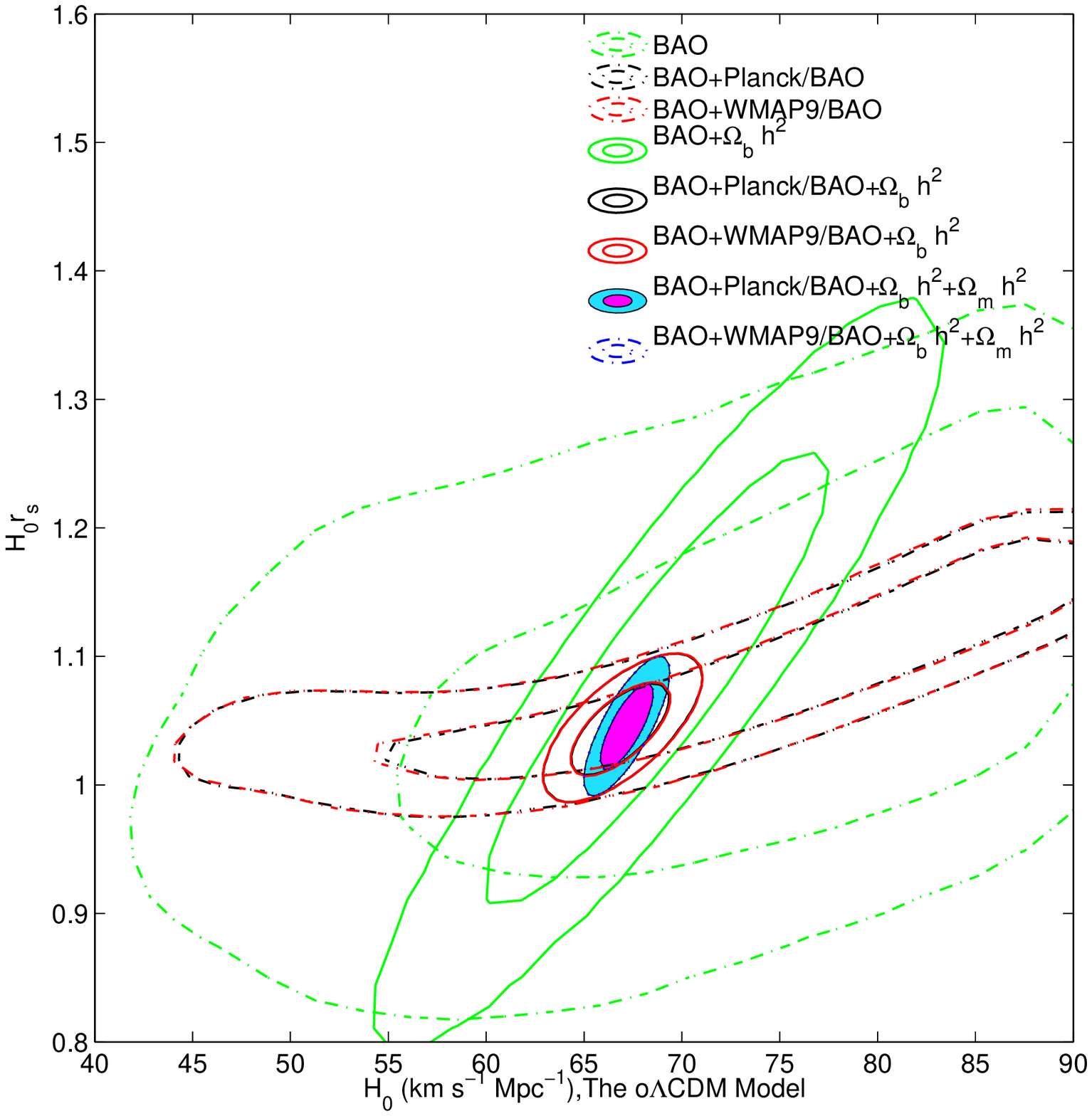}}\quad
            {\includegraphics[width=2.0in]{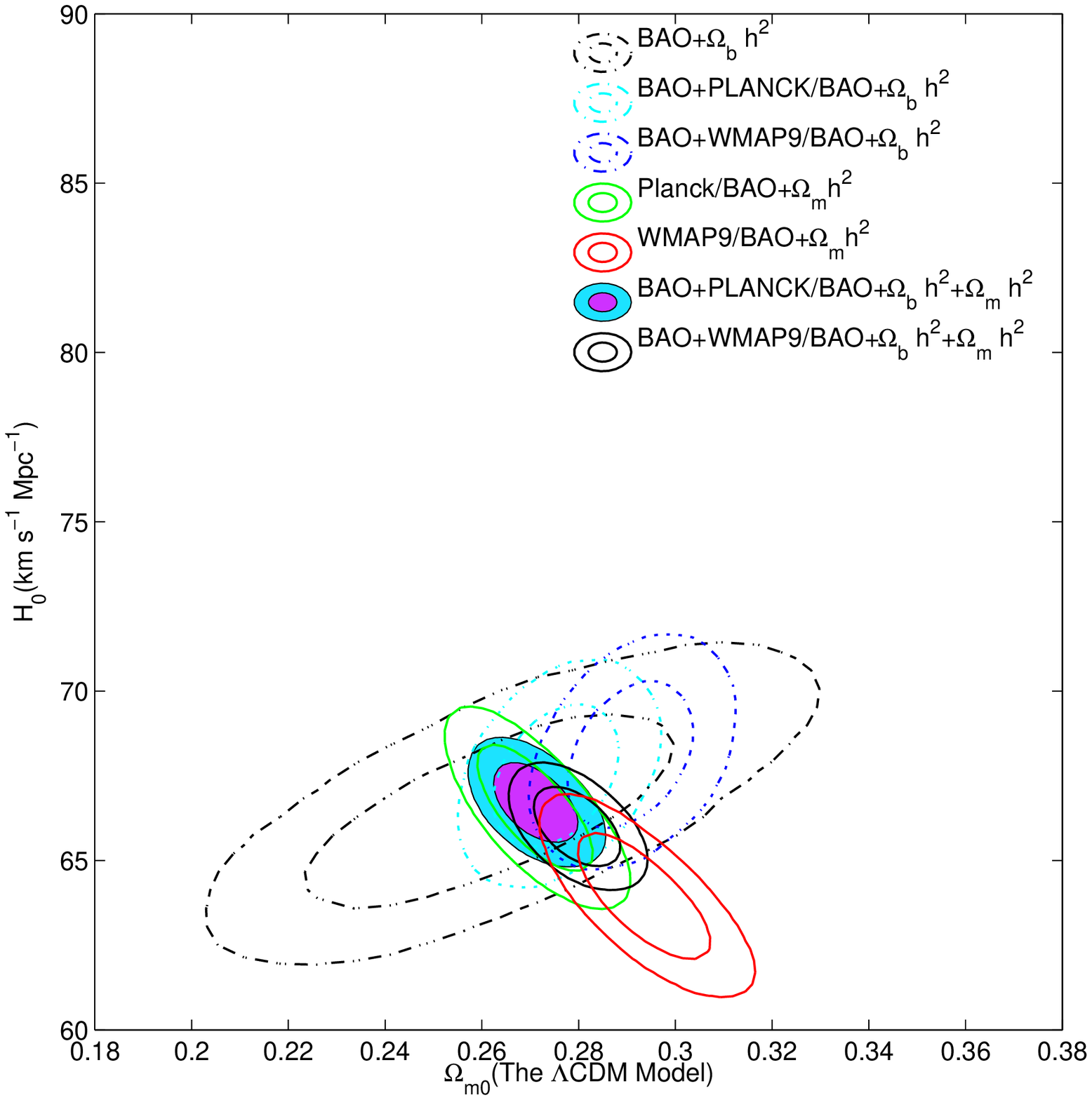}}\quad
             {\includegraphics[width=2.0in]{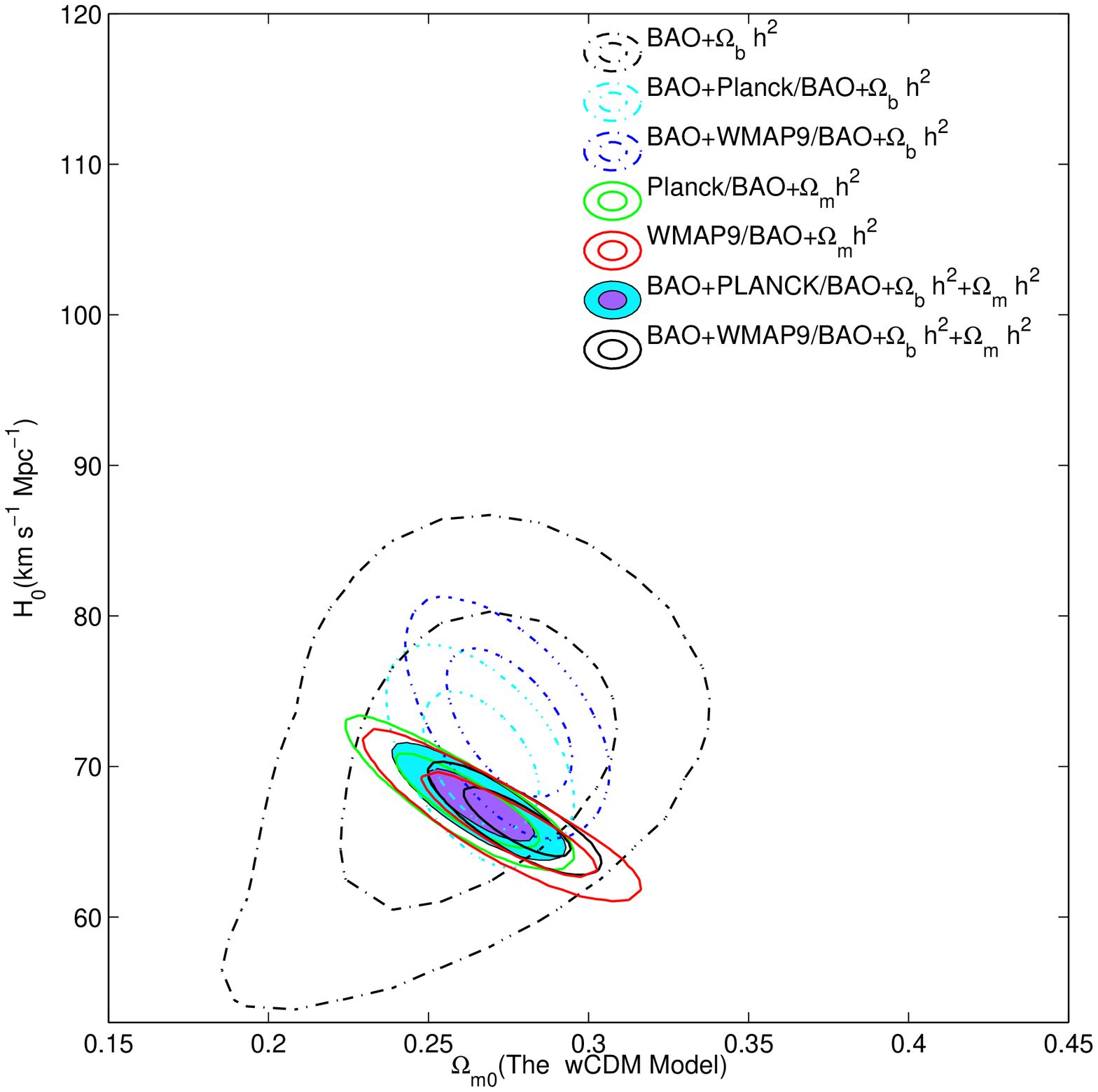}}\quad
                  {\includegraphics[width=2.0in]{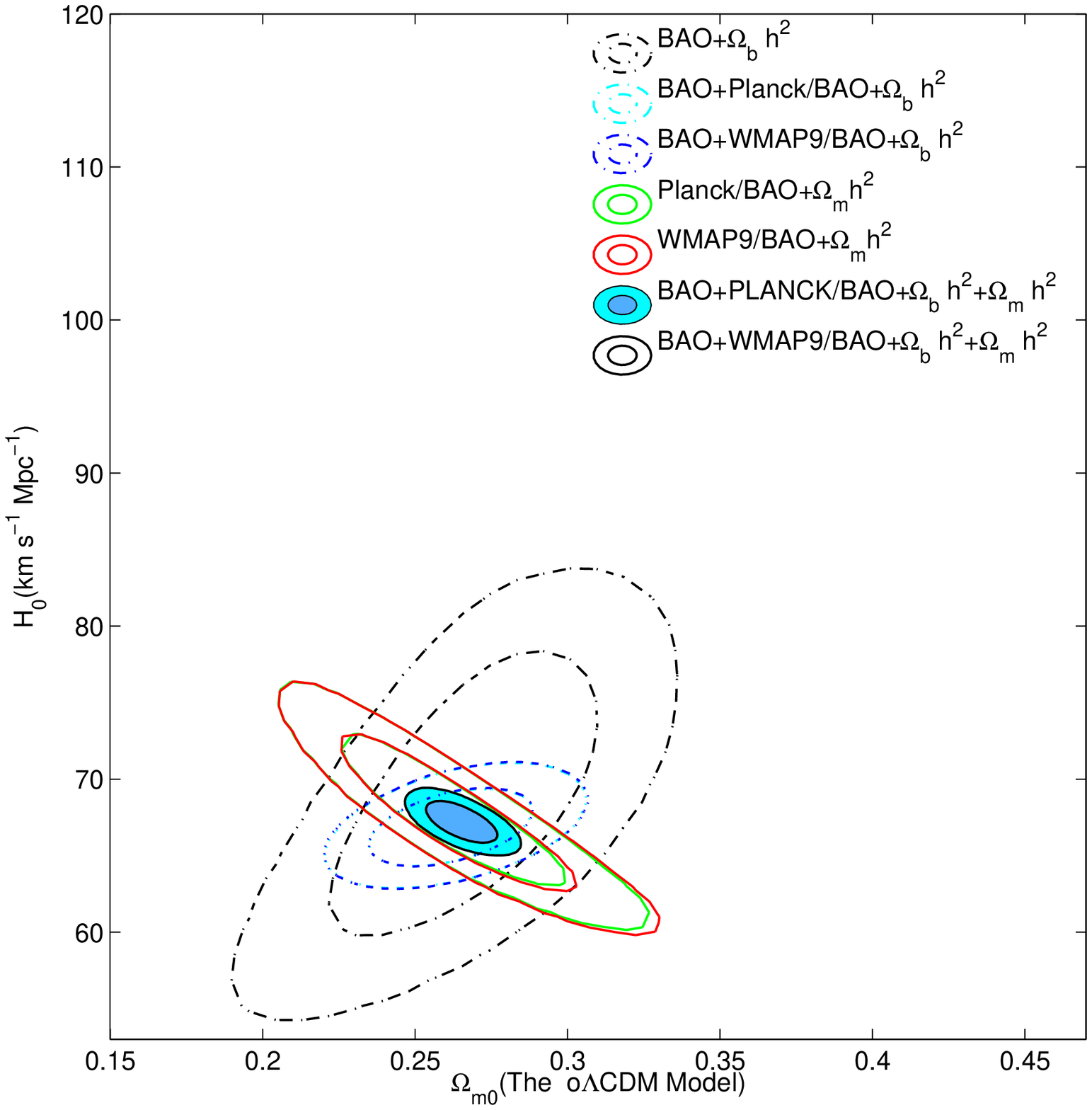}}\quad
    \end{center}
 \caption{\small  The upper and lower three panels are  the contour plots of  $H_0r_s-H_0$ and  $H_0-\Omega_{m0}$ separatly  for the $\Lambda$ CDM, $w$CDM and o$\Lambda$ CDM models.   }\label{h}
\end{figure}

\section{Short Summary}\label{sec4}

Here we obtain  CMB/BAO samples with 13  data in the range of $0.106 \leq z \leq 2.34$ and focus  on parameter constraints and model tests.   
Basically, the CMB/BAO data give out a tighter constraint compared to the BAO data though its error increased.  
As the degeneracies of $\Omega_{m0}-w$  and  $\Omega_{m0}-\Omega_{k0}$  are positive for the CMB/BAO data while it is negative for the BAO distance ratio data.
 
Fitting the theoretic  models to  the BAO+Planck/BAO+$\Omega_bh^2$+$\Omega_mh^2$ data , we get constraints on $\Lambda$CDM  model as $\Omega_{m0}= 0.271_{-0.010-0.015}^{+ 0.011+ 0.016}$ and $H_0=66.7_{ -1.4 -2.1}^{+  1.4+  2.1}\,km\, s^{-1} \,Mpc^{-1}$ ;   constraints on the w$\Lambda$CDM model  as $\Omega_{m0}=0.267_{-0.024-0.035}^{+ 0.022+ 0.031}$, $w=-1.042_{-0.242-0.357}^{+ 0.196+ 0.267}$ and $H_0= 67.3_{ -3.0 -4.2}^{+  3.6+  5.4} \,km\, s^{-1} \,Mpc^{-1}$ and  constraints on the o$\Lambda$CDM model from the Planck/CMB as $\Omega_{m0}=0.265_{-0.016-0.022}^{+ 0.017+ 0.024}$,  $\Omega_{k0}=-0.017_{-0.013-0.018}^{+ 0.014+ 0.020}$ and $ H_0=67.2_{ -1.9 -2.6}^{+  1.9+  2.7}\,km\, s^{-1} \,Mpc^{-1}$.

 All our data about $\Omega_{m0}$ have tension with the Planck result, but consistent with  the SNe data. 
 As for  $H_0$, our result is consistent with the Planck data, but has tension with the local measurements.
And, our results slightly favor  phantom dark energy  where $w<-1$ and a negative $\Omega_{k0}$.


\section{Acknowledgements}
The author thank  Dr. Hui Li  and  Dr. Zhengxiang Li for useful discussions.
This work was supported by    the National Natural Science
Foundation of China key project under grant No. 10935013, and
CQ CSTC under grant
No. 2010BB0408.


{}
\end{document}